\documentstyle[12pt]{article}
\newcommand{\al}{\alpha'}
\newcommand{\de}{\partial}
\newcommand{\be}{\begin{equation}}
\newcommand{\ba}{\begin{eqnarray}}
\newcommand{\ea}{\end{eqnarray}}
\newcommand{\ee}{\end{equation}}

\newcommand{\we}{\wedge}
\newcommand{\ca}{\mathcal}
\newcommand{\lr}{\leftrightarrow}
\newcommand{\f}{\frac}
\newcommand{\s}{\sqrt}
\newcommand{\ti}{\tilde}
\newcommand{\ap}{\alpha}

\newcommand{\mb}{\mathbf}
\newcommand{\ddd}{\cdot\cdot\cdot}
\newcommand{\dd}{\cdot\cdot}
\newcommand{\no}{\nonumber \\}

\newcommand{\ep}{\epsilon}
\newcommand{\zz}{{\mathbf Z}_2}
\newcommand{\z}{{\mathbf Z }}
\newcommand{\io}{{\mathbf i }}
\newcommand{\beq}{\begin{equation}}
\newcommand{\eeq}{\end{equation}}
\newcommand{\beqa}{\begin{eqnarray}}
\newcommand{\eeqa}{\end{eqnarray}}
\newcommand{\CR}{\nonumber \\}
\newcommand{\pa}{\partial}
\newcommand{\A}{\alpha}

\newcommand{\p}{\Phi}

\begin{document}
\begin{titlepage}
\thispagestyle{empty}
\begin{flushright}
UT-919 \\
hep-th/0012210 \\
December, 2000 
\end{flushright}

\bigskip
\bigskip

\begin{center}
\noindent{\Large \textbf{Brane-Antibrane Action\\
\bigskip
from Boundary String Field Theory}}\\
\vspace{2cm}
\noindent{
          Tadashi Takayanagi\footnote{takayana@hep-th.phys.s.u-tokyo.ac.jp}
, Seiji Terashima\footnote{seiji@hep-th.phys.s.u-tokyo.ac.jp} 
and Tadaoki Uesugi\footnote{uesugi@hep-th.phys.s.u-tokyo.ac.jp} }\\

\bigskip
{\it Department of Physics, Faculty of Science \\ University of Tokyo \\
\medskip
Tokyo 113-0033, Japan}

\end{center}
\begin{abstract}
In this paper we give the boundary string field theory description of  
brane-antibrane systems. From the world-sheet action of brane-antibrane 
systems we obtain the tachyon potential and discuss the tachyon 
condensation exactly. We also find the world-volume action including the 
gauge fields. Moreover we determine RR-couplings exactly 
for non-BPS branes and brane-antibranes. These couplings 
are written by superconnections and correspond to $K^1(M)$ and 
$K^0(M)$ for the non-BPS branes and brane-antibranes, respectively. 
We also show that Myers terms appear if we include the transverse scalars 
in the boundary sigma model action. 
\end{abstract}
\end{titlepage}

\newpage

%\tableofcontents

\section{Introduction}

In recent years there has been a lot of work on tachyon physics \cite{sen18}. 
In string theory tachyon fields naturally appear on the non-BPS branes 
\cite{BeGa1, sen14, sen16, senI} and 
the brane-antibrane systems \cite{Gr, BaSu, sen13}. 
These studies of the dynamical aspects of non-BPS systems are very 
important to understand the vacuum structure of 
open string theory because we can 
always see 
the process of the vacuum 
transition from an unstable one to a stable one via the tachyon
condensation. 
In the special case 
this process can be analyzed via the marginal deformation of conformal field 
theory \cite{sen14, sen16, FrGaLeSt, MaSe2, NaTaUe}. 
However in the general situation the tachyon 
condensation is an off shell phenomenon. 
Thus we should use string field theories. 

Historically the most famous string field theory-Witten's cubic string field 
theory- has been mainly used to compute the tachyon potential of bosonic 
branes, non BPS branes and brane-antibranes by the approximation which is 
called the level truncation (for example see \cite{cubic}).
This approximation is needed because generally in the process of the tachyon 
condensation many higher massive modes on a D-brane are excited, 
and we can not consider the infinite numbers of modes at the same time. 

However quite recently 
another string field theory has been applied to the tachyon 
condensation. Some exact tachyon potentials and effective actions including 
both the tachyon and gauge fields were calculated \cite{GeSh1, KuMaMo1, 
KuMaMo2, MoNa, Da, FuIt}. 
This is called background independent open string field theory (BIOSFT) 
or boundary string field theory (BSFT), which was first formulated by 
Witten \cite{Wi2}. 
This string field theory is based on the world-sheet sigma model 
action which is perturbed by the relevant operators on the boundary of disk. 
The strategy for the exact analysis is the following. 
If we put the profile of the 
tachyon field to the special form which makes the world-sheet sigma model action 
become free, then the massive modes on 
branes are not excited due to the renormalization of the world-sheet
theory \cite{GeSh1, KuMaMo1}. 
Therefore the calculations including only the tachyon field are exact. 

The boundary string field theory was first formulated for bosonic 
open string. Its superstring version is not known. However from the 
argument of the world-sheet supersymmetry and 
the boundary entropy the authors of \cite{KuMaMo2} conjectured 
that the string field theory action is equal to 
the partition function. If this conjecture is correct, we can calculate more
 easily the string field theory action than in the bosonic case. 
In \cite{KuMaMo2} by using this conjecture the effective action of
non-BPS brane 
was calculated and this result was equal to the proposed form in several 
papers \cite{sen22, MiZw, Ga, BeRoWiEyPa} if we assume that 
the tachyon field is constant. 

On the other hand only limited results for the effective action of the 
brane-antibrane have been obtained. For the results from on-shell scattering 
amplitude, see \cite{pe}. However the explicit form of the effective action 
is very important to know the dynamical aspects of the tachyon condensation 
in the brane-antibrane systems. For example, if we want to discuss the 
noncommutative tachyon on the brane-antibrane, then the detailed form of the 
effective action is required \cite{HaKrLa1}. These dynamical aspects of the 
brane-antibrane system such as its effective action can only be obtained by 
the off-shell calculations. Especially the boundary string field theory is 
suitable for investigating the general structure of the effective action 
exactly.

Therefore one of the purpose of this paper is to study the effective action 
of the brane-antibrane system in boundary string field theory.
Indeed, we obtain {\it exact} results for the tachyon condensation.
Especially we show that 
by considering the special profile of tachyons (kink or vortex) lower 
dimensional D-branes are produced and these tensions are equal to the known 
values exactly. We also discuss the general non-abelian cases and show that 
so called Atiyah-Bott-Shapiro (ABS) construction \cite{ABS} 
naturally appears in 
the boundary string field theory.

However if we include the gauge field, the tachyon on 
the brane-antibrane couples to two kinds of the gauge fields in the 
bi-fundamental representation. Therefore there is no choice of the profiles of 
the tachyon and the gauge field to make the world-sheet action free.
And it is difficult to obtain the exact effective action 
for the tachyons and gauge fields. 
This is different from abelian non-BPS case \cite{KuMaMo2} 
(in this case the tachyon does not couple to the gauge field because the 
tachyon is in the U(1) adjoint representation). 
However it is possible to calculate 
several lower terms in the $\alpha^{\prime}$ expansion. And we can also 
discuss the general form of the effective action. 
Indeed this general form is consistent 
with the argument on the noncommutative solitons \cite{HaKrLa1}. 
 
Above arguments are limited to the effective action for NSNS sector. 
However the boundary 
string field theory has the remarkable property that the on-shell RR closed 
vertex can be inserted, while it is difficult for the cubic string field 
theory. Moreover in the 
boundary string field theory we can formally incorporate the gauge
fields at any order, while in the cubic theory we can introduce the
gauge fields only by the perturbation. 
If 
one notices that 
the boundary interactions for the gauge field strengthes 
are similar to
the RR-couplings of BPS D-branes 
represented by the Chern character,
one expects that in the boundary string field theory RR-couplings of the 
non-BPS branes and brane-antibranes is computable {\it exactly}. 
Indeed this expectation is true. Therefore in this paper we give the most 
general coupling forms in the case of non-BPS branes and of 
brane-antibrane systems. These are represented by the so called 
{\it superconnection} \cite{Qu}, which was conjectured in \cite{Wi1, KeWi} 
in the case of brane-antibranes. 
We can also show that the RR coupling of non-BPS branes 
has the structure of the superconnection. In mathematics it is known
that the charge which is represented by Chern character of a
superconnection is equivalent to K-theory charge. This means that the 
RR-coupling of brane-antibranes and non-BPS branes 
corresponds to $K^0(M)$ and $K^1(M)$ respectively. 
Therefore this gives another evidence of K-theory classification of
D-brane charges \cite{MiMo,Wi1,Ho1}.

The plan of the paper is the following.
   
In section 2 we review the boundary string field theory for non-BPS branes and 
present the world-sheet action for the brane-antibrane system. 
We justify this world-sheet action by showing that with putting the tachyon 
field to zero the partition function becomes the sum of two DBI 
actions, which is one for a brane and the other for an antibrane.  

In section 3 we study exact tachyon condensations for special profiles of 
the tachyon and show that the tensions of lower D-branes which 
are produced after the tachyon condensation are equal to known values in 
general situations and we relate these tachyon profiles to Atiyah-Bott-Shapiro 
construction.

In section 4 we calculate RR couplings for brane-antibranes and non-BPS 
D-branes by boundary string field theory. We also discuss that these forms are 
written by the superconnections, and we relate these to K-theory groups.
In the last subsections we generalize these couplings to the couplings 
including noncommutative transverse scalars, which is called Myers term \cite{My}. 
 
In section 5 we calculate the effective action for NSNS sector. 
We show that the form of the action is the sum of the DBI actions
multiplied by the tachyon potential and 
that this action is consistent with the argument of the noncommutative 
soliton. We also calculate several lower terms in the $\alpha^{\prime}$ 
expansion.

In appendix we summarize the notations and spinor formulas 
mainly for the calculation of RR-couplings.

\section{World-Sheet Action of Brane-Antibrane System}

Recently using background independent open string field theory several 
effective actions has been calculated exactly 
in a certain sense \cite{GeSh1, 
KuMaMo1, KuMaMo2}.
 This string field theory (from now on we call this BSFT) 
was first considered by Witten in the bosonic open string 
field theory \cite{Wi2}. 
In that paper the open string field action was defined by extending 
Batalin-Vilkovisky (BV) formalism to that for open-string fields. 
The solution of this string field master equation was given by \cite{Sh1}:
\begin{eqnarray}
\label{eq11}
S=(\beta^i(\lambda)\frac{\partial}{\partial\lambda^i}+1)Z,
\end{eqnarray}
where $S$ is the string field action, $Z$ is the partition function, 
$\lambda^i$ is one dimensional coupling 
of sigma model (i.e. target space field) and 
$\beta^i(\lambda)$ is beta function of it.

This is for bosonic open string field theory. 
The BV-like formulation of background independent superstring field theory 
has not been found until now. However the relation between {\sl S} and $Z$ 
in (\ref{eq11}) can be generalized to the supersymmetric version. 
Some years ago Tseytlin et.al.\cite{AnTs1, Ts1} 
calculated several partition functions including only the gauge fields 
and they confirmed that partition functions were equal to the effective 
actions constructed by calculating S-matrix perturbatively in supersymmetric 
case (not in bosonic case). Moreover they conjectured that this partition 
function can be identified with {\it off shell} string field action. 
In \cite{KuMaMo2} they extended these interpretations to the full open string 
field theory including tachyons.

Therefore in this paper we expect that the same relation holds not only for 
non-BPS branes but also for brane-antibrane systems:
\begin{eqnarray}
\label{eq4}
S &=& Z .
\end{eqnarray}

Below we propose a brane-antibrane sigma model action and 
calculate the string field action. This sigma model action is 
the extension of non-BPS brane's one, thus before giving this  
we first review non-BPS brane's one\cite{KuMaMo2,Ts2}.  

The partition function is defined by:
\begin{eqnarray}
\label{eq5}
Z &=& \int DX D\psi D\eta \exp[-I(X,\psi,\eta)].
\end{eqnarray}

In this definition the action of the $\sigma$ model $I$ is: 
\begin{eqnarray}
I &=& I_0+I_B,\\
I_0 &=& \frac{1}{4\pi}\int_{\Sigma}d^2 z [\partial_z X^{\mu} 
\partial_{\bar{z}}X_{\mu}+\psi^{\mu}\partial_{\bar{z}}\psi_{\mu}
+\tilde{\psi}^{\mu}\partial_{z}\tilde{\psi}_{\mu}],\\
\label{eq1}
I_B &=& \int_{\partial\Sigma}d\tau d\theta[-{\bf \Gamma}D_{\theta}{\bf \Gamma}
+\frac{1}{\sqrt{2\pi}}T({\bf X}){\bf \Gamma}
-iD_{\theta}{\bf X}^{\mu}A_{\mu}({\bf X})].
\end{eqnarray}
where the superspace representation in the boundary theory is defined by:
\footnote{In this paper we set $\alpha^{\prime}$ to 2.}
\begin{eqnarray}
\left\{\begin{array}{lcl}
{\bf X}^{\mu} &=& X^{\mu}+2i\theta\psi^{\mu},\\
{\bf \Gamma} &=& \eta+\theta F,\\
D_{\theta} &=& \frac{\partial}{\partial\theta}
+\theta\frac{\partial}{\partial\tau}.\end{array}\right.
\end{eqnarray}

If one writes $I_B$ in the component form 
and integrate out the auxiliary field $F$,\\
 then it becomes:
\begin{eqnarray}
I_B = \int_{\partial\Sigma}d\tau [\frac{1}{8\pi}T(X)^2+\eta\dot{\eta}
+i\sqrt{\frac{2}{\pi}}\psi^{\mu}\eta\partial_{\mu}T-i\dot{X}^{\mu}A_{\mu}
+2iF_{\mu\nu}\psi^{\mu}\psi^{\nu}]. \label{bin}
\end{eqnarray}

This is the world-sheet action for a non-BPS brane.\\
The superfield ${\bf \Gamma}$ corresponds to the internal 
degrees of the freedom of non-BPS branes which is equal to 2$\times$2 
matrices $1,\sigma_2$ (Pauli matrix) \cite{senI, sen18}. This ${\bf \Gamma}$ 
field description is first given by Witten \cite{Wi1} and 
Harvey et.al. proposed that this action (\ref{bin}) describes 
non-BPS branes in \cite{HaKuMa}.

The tachyon field $T({\bf X})$ is gauge-transformed in the U(1) adjoint 
representation (that is equal to the gauge singlet), thus this action is gauge 
invariant without ${\bf \Gamma}$ being gauge transformed.
This makes $T({\bf X}),{\bf \Gamma}$ decoupled from the gauge field,
which fact appears in the (\ref{eq1}). Therefore if one exponentializes
the action $I$ and performs the path-integration 
in the approximation of neglecting the third term of 
(\ref{bin}), the partition function $Z$(=S) becomes the simple structure which 
is the product of the DBI action and tachyon potential 
$\exp(-\frac{1}{4}T^2)$ \cite{sen22, Ga, BeRoWiEyPa, KuMaMo1, KuMaMo2}.

However if we consider non-abelian non-BPS D-branes, the tachyon field 
couples to the gauge fields, its action is more complicated than U(1) case 
and  the calculation of effective action is difficult. Non-abelian
action was proposed in \cite{Ts2}:
\begin{eqnarray}
I_B &=& \int_{\partial\Sigma}d\tau d\theta [-{\bf \bar{\Delta}}D_{\theta}
{\bf \Delta}-{\bf \Gamma}D_{\theta}{\bf \Gamma}
+{\bf \bar{\Delta}}\{-\f{1}{\s{2\pi}}{\bf \Gamma}
T({\bf X})+iA_{\mu}({\bf X})D{\bf X}^{\mu}\}
{\bf \Delta}],  \label{nb-n}
\end{eqnarray} 
where T({\bf X}) is $N\times N$ matrix in the case of N non-BPS D-branes, 
${\bf \Delta}$ is complex fermionic superfield which couples to the 
gauge field in the fundamental representation.

Next, we want to extend this action to the brane-antibrane system
\footnote{In this paper we mainly consider one pair brane-antibrane case. 
About the generic configuration of brane-antibrane system (m D-brane 
+ n anti D-brane) we comment at several points.}, which contains tachyons 
and is unstable.
Before considering the world-sheet action, we should be reminded of the 
characteristic properties of this system (For a review, see \cite{sen18}).
First, a brane-antibrane system has two kinds of vector multiplets
(gauge fields and GSO even fermions). One lives in D-brane, 
another in anti D-brane.
Second, the tachyons and GSO odd fermions come
from the open strings between a D-brane and an anti D-brane.
In Type II theories open strings have the
orientation, thus the D-brane and the anti D-brane 
have two kinds of real tachyon fields and 
we can represent these by a complex tachyon field 
which belongs to the bi-fundamental representation.
Third, this system is essentially non-abelian(this contains at least 
two branes in one pair system) and this contains the 
Chan-Paton factors. Especially in the one pair case, the Chan-Paton factors
are represented by $2\times2$ matrices (identity matrix and Pauli matrices
($\sigma_1,\sigma_2,\sigma_3$)). The identity matrix represents the freedom 
of the total sum of gauge fields of the system.
The matrix $\sigma_3$ represents the freedom of the 
relative difference of gauge fields and $\sigma_1,\sigma_2$ the ones of 
tachyons. These are the main properties of brane-antibrane systems.

It is famous that these are related to non-BPS D-branes by the ``descent 
relation'' conjectured by Sen \cite{sen18,sen16,sen27}. 
Therefore we expect that the world-sheet action is very similar to the one of 
non-BPS branes. Naively the real tachyon field $T(X)$ in (\ref{eq1}) 
is extended to the complex field $T({\bf X})$, $\bar{T}({\bf X})$ which 
is gauge-transformed as follows:
\begin{eqnarray}
\label{eq2}
T({\bf X})\rightarrow e^{i\lambda_1({\bf X})}T({\bf X})
e^{-i\lambda_2({\bf X})},
\end{eqnarray}
where $\lambda_1, \lambda_2$ are arbitrary functions of ${\bf X}$. 
The sigma model action for the non-BPS D-brane (\ref{eq1}) 
respects gauge-symmetry\footnote{Strictly 
speaking this symmetry is not two dimensional gauge symmetry but non-linear 
global symmetry. However it is famous in usual sigma model(Type I, Heterotic)
that world-sheet global symmetry corresponds to target space gauge symmetry} 
and world-sheet supersymmetry, thus in the case of the brane-antibrane
it is natural to require these symmetries.  
{}From these considerations we propose that the following action 
defines the $D9-\overline{D9}$ action in BSFT (One for $Dp-\overline{Dp}$ 
is simply obtained by T-duality).
\begin{eqnarray}
I &=& I_0+I_B,\\
I_0 &=& \frac{1}{4\pi}\int_{\Sigma}d^2 z[\partial_z X^{\mu}
\partial_{\bar{z}}X_{\mu}+\psi^{\mu}\partial_{\bar{z}}\psi_{\mu}
+\tilde{\psi}^{\mu}\partial_{z}\tilde{\psi}_{\mu}],\\
\label{eq10}
I_B &=& \int_{\partial\Sigma}d\tau d\theta[-{\bf \bar{\Gamma}}(D_{\theta}
-iA_{\mu}^{(-)}({\bf X})D_{\theta}{\bf X}^{\mu}){\bf \Gamma}
+\frac{1}{\sqrt{2\pi}}{\bf \bar{\Gamma}}T({\bf X})\nonumber\\
\label{eq3}
&&+\frac{1}{\sqrt{2\pi}}\bar{T}({\bf X}){\bf \Gamma}
-\frac{i}{2}D_{\theta}{\bf X}^{\mu}A^{(+)}_{\mu}({\bf X})].
\end{eqnarray}

If we write $I_B$ in the component form and integrate out 
the auxiliary fields $F,\bar{F}$ in ${\bf \Gamma}$ and ${\bf \bar{\Gamma}}$: 
\begin{eqnarray}
\label{eq6}
I_B &=& \int_{\de\Sigma} d\tau [\bar{\eta}\dot{\eta}+2i\bar{\eta}\eta\psi^{\mu}
\psi{^\nu}F^{(-)}_{\mu\nu}-i\bar{\eta}\eta \dot{X^\mu}A^{(-)}_{\mu} \no
& &-i\sqrt{\frac{2}{\pi}}\bar{\eta}\psi^{\mu}D_{\mu}T
+i\sqrt{\frac{2}{\pi}}\psi^{\mu}\eta \overline{D_{\mu}T}
+\frac{1}{2\pi}\bar{T}T-\f{i}{2}\dot{X}^{\mu}A^{(+)}_{\mu}+i\psi^{\mu}
\psi^{\nu}F^{(+)}_{\mu\nu}],
\end{eqnarray}
where we have employed the following definition:
\begin{eqnarray}
\left\{\begin{array}{lcl}
A^{(\pm)}_{\mu} &=& A^{(1)}_{\mu}\pm A^{(2)}_{\mu},\\
D_{\mu}T &=& \partial_{\mu}T-iA^{(-)}_{\mu}T,\\
F^{(1),(2)}_{\mu\nu} &=& \partial_{\mu}A^{(1),(2)}_{\nu}
-\partial_{\nu}A^{(1),(2)}_{\mu}. \end{array} \right.
\end{eqnarray}

The field ${\bf \Gamma}$ is gauge-transformed in the bi-fundamental 
representation which is same as (\ref{eq2}). 
This fact forces the first term in
 (\ref{eq3}) to be gauged.
This is crucial difference from the non-BPS brane case
 (\ref{eq1}). This prescription is usual in Type I or Heterotic non-linear 
sigma model action (for example see the section 12.3 in \cite{Po1}) .

On the other hand, if we construct the most general 
one-dimensional {\sl renormalizable} action
which is written by superfields ${\bf \Gamma}, {\bf \bar{\Gamma}}$ and
${\bf X}$, then this action coincides with (\ref{eq3}) up to the
arbitrary real function $g({\bf X})$ in front of the first term of
(\ref{eq3}). This may be confusing, because it looks as if the new real
scalar field $g({\bf X})$ appeared on the
brane-antibrane. However, this 
$g({\bf X})$ can be eliminated by the redefinition of ${\bf \Gamma}$, 
$T({\bf X})$ and $A^{\pm}_{\mu}({\bf X})$. Therefore this $g({\bf X})$
is a redundant two-dimensional sigma model coupling, and even if we set
$g({\bf X})$ to 1, the renormalizability is respected. At any rate 
eq.(\ref{eq3}) is the renormalizable action. We will use this fact in
section 5.

{}From this action, we can calculate the effective action of the 
brane-antibrane system. Especially we are interested in the form of 
NSNS effective action of the brane-antibrane system, 
which corresponds to the Dirac-Born-Infeld (DBI)
part of BPS D-brane action,
because the explicit form of this 
effective action is not known as much as non-BPS one
(In the non-BPS case the action is more familiar 
than the brane-antibrane. 
This is proposed in several papers \cite{sen22,Ga,BeRoWiEyPa}).   

Since above arguments of constructing brane-antibrane action are
too heuristic, we have to confirm that this action describes the 
brane-antibrane system correctly from several point of view. 
In this paper, we confirm three nontrivial checks before calculating 
the full NSNS action of the brane-antibrane as follows:
\begin{itemize}
\item{With setting $T(X)$ to 0 in (\ref{eq3}) we reproduce the sum of the DBI
 action of two kinds of gauge fields.}
\item{By considering the tachyon condensation 
without the gauge fields by BSFT we check the descent 
relation between the non-BPS D-brane and the brane-antibrane.} 
\item{We reproduce {\it exactly} the RR couplings of the brane-antibranes  
conjectured in \cite{KeWi} which are represented by superconnection formula 
\cite{Qu}.}
\end{itemize}

At first sight the first fact looks false because even if we set $T(X)=0$, 
it is likely that massive modes which fly between a D-brane and an antiD-brane 
modify the sum of DBI actions. However this effect comes from open string one 
loop effect (cylinder amplitude) and  in the disk amplitude 
this effect does not cause any modification. 

Now we check the first fact.
The second and the third fact will be checked in the section 3 and 4, 
respectively, and finally in the section 5 we calculate the NSNS action. 
The path integral representation 
of the partition function is given by (\ref{eq5}).
First we split $Z$ into two dimensional part (internal of disk) and one 
dimensional part (boundary of the disk):
\begin{eqnarray}
\label{eq7}
Z=\int DX D\psi \exp[-I_0(X,\psi)] 
\int D\eta D\bar{\eta} \exp[-I_B(X,\psi,\eta,\bar{\eta})].  
\end{eqnarray} 

In open string NS sector $\psi(\tau)$ obeys the anti-periodic boundary 
condition so that $\psi(\tau)$ has half-integer Fourier modes. 
Then $\eta(\tau)$ and $\bar{\eta}(\tau)$ should also 
obey the anti-periodic boundary condition 
in order for (\ref{eq6}) to be locally well defined. 

Here we integrate $\eta(\tau)$ and $\bar{\eta}(\tau)$ first. 
The path integral of $\eta(\tau)$ and $\bar{\eta}(\tau)$ 
is defined on the circle, 
which corresponds to one loop partition function. 
Therefore transforming this path integral to Hamiltonian formalism, 
$\eta$ and $\bar{\eta}$ are quantized and from (\ref{eq6})
these obey the usual canonical quantization condition:
\begin{eqnarray}
\{\eta,\bar{\eta}\}=1.
\end{eqnarray}

By using canonical quantization method of $\eta$ and $\bar{\eta}$,
the partition function $Z$ becomes as follows:
\begin{eqnarray}
\label{eq8}
Z &=&\int DX D\psi \exp[-I_0(X,\psi)]\nonumber\\ 
\!\! &\times& \!\!
{\rm Tr~P}\exp\int^{\pi}_{-\pi}d\tau\biggl[i\frac{[\bar{\eta},\eta]}{2}
\dot{X}^{\mu}A^{(-)}_{\mu}(X)-2i\frac{[\bar{\eta},\eta]}{2}\psi^{\mu}\psi^{\nu}
F^{(-)}_{\mu\nu}(X)+i\sqrt{\frac{2}{\pi}}\bar{\eta}\psi^{\mu}D_{\mu}T(X)
\nonumber\\
&&~~-i\sqrt{\frac{2}{\pi}}\psi^{\mu}\eta\overline{D_{\mu}T(X)}
-\frac{1}{2\pi}\bar{T}(X)T(X)+\frac{i}{2}\dot{X}^{\mu}A^{(+)}_{\mu}(X)
-i\psi^{\mu}\psi^{\nu}F^{(+)}_{\mu\nu}(X)\biggr],
\end{eqnarray}
where ``${\rm P}$'' represents the path ordering and 
``${\rm Tr}$''(trace) implies that we should sum expectation values 
in two state Hilbert space:
\begin{eqnarray}
\eta|\downarrow\rangle=0~&,&~\bar{\eta}|\downarrow\rangle=|\uparrow\rangle,
\nonumber\\
\eta|\uparrow\rangle=|\downarrow\rangle~&,&~\bar{\eta}|\uparrow\rangle=0.
\end{eqnarray}

When we construct Hamiltonian from Lagrangian, 
we set the operator ordering by antisymmetrization of $\eta$
and $\bar{\eta}$.

This is a consequence from the quantum mechanics, but at a first glance it is 
strange. We said that classical fields $\eta, \bar{\eta}$ obey 
anti-periodic boundary conditions and have half-integer modes.
Therefore they do not have zero-modes. However if we use Hamiltonian 
formalism, $\tau$ dependence of $\eta, \bar{\eta}$ drops out and it looks 
like that only zero modes remain. 
This is confusing. Yet if we perform the path 
integral simply by the perturbation using $\eta, \bar{\eta}$ Green function 
with anti-periodic boundary condition:
\begin{eqnarray}
\langle\eta(\tau)\bar{\eta}(\tau^{\prime})\rangle
&=&\frac{1}{2}\epsilon(\tau-\tau^{\prime})
=\frac{1}{\pi}\sum_{r \in \z +\frac{1}{2}>0} \frac{{\rm sin\{r(\tau-\tau')\}}}{r},\\
&&\epsilon(\tau)\equiv
\left\{\begin{array}{c}
1~(\tau>0)\\
0~(\tau=0)\\
-1~(\tau<0) \end{array}\right. ,
\end{eqnarray}
then we can check order by order 
that path integral representation (\ref{eq7}) is equal to (\ref{eq8}).
However note that this identity holds only for 
$\epsilon(0)=0$ regularization.

Here we can check that the operator commutation and anti-commutation 
relation of $\bar{\eta}, \eta ,[\bar{\eta},\eta]$ are same as that of 
Pauli matrix, $\sigma_{+}, \sigma_{-}, \sigma_3$ (where $\sigma_{\pm}\equiv 
\frac{1}{2}(\sigma_1\pm i\sigma_2)$). 
Therefore we can replace 
$\bar{\eta}, \eta, [\bar{\eta},\eta]$ by $\sigma_{+}, \sigma_{-},
\sigma_3$, respectively.
   
In that form $Z$ becomes:
\begin{eqnarray}
\label{eq9}
& &Z~=~\int DX D\psi \exp[-I_0(X,\psi)]\times{\rm Tr~P}\exp\int^{\pi}_{-\pi} 
d\tau M(\tau)~,
\end{eqnarray}
where
\begin{eqnarray}
& &M(\tau)~=~\left(\begin{array}{cc}
i\dot{X}^{\mu}A^{(1)}_{\mu}-2i\psi^{\mu}\psi^{\nu}F^{(1)}_{\mu\nu}
-\frac{1}{2\pi}T\bar{T}~&~ i\sqrt{\frac{2}{\pi}}\psi^{\mu}D_{\mu}T
\nonumber\\
-i\sqrt{\frac{2}{\pi}}\psi^{\mu}\overline{D_{\mu}T}~&~i\dot{X}^{\mu}
A^{(2)}_{\mu}-2i\psi^{\mu}\psi^{\nu}F^{(2)}_{\mu\nu}
-\frac{1}{2\pi} \bar{T}T \nonumber
\end{array}\right).\\
\label{add}
\end{eqnarray}

This is one expression of the brane-antibrane partition function.
Furthermore this form is able to be extended to the generic
configuration of the 
brane-antibrane system ($m$ D-branes and $n$ antiD-branes) 
because in that case we have only to replace 
$A^{(1)}_{\mu}(X),A^{(2)}_{\mu}(X)$ and $T(X)$ in (\ref{add}) with 
$m\times m, n\times n$ and $m\times n$ matrices, respectively, 
while the expression (\ref{eq9}) has one fault that the gauge symmetry and 
the world-sheet supersymmetry cannot be seen explicitly. We could not find 
(\ref{eq10}) type action in the case of the non-abelian brane-antibrane 
system.
   
Then we go back to the original question. That is 
``With setting $T(X)$ to 0 in (\ref{eq3}) can we reproduce 
the sum of the DBI actions ?'' 
If one sets $T(X)$ to 0 in (\ref{add}), the off-diagonal part vanishes and 
the diagonal part remains. In this case $Z$ becomes as follows:
\begin{eqnarray}
Z &=&\int DX D\psi \exp[-I_0(X,\psi)]\nonumber\\ 
&\times&{\rm Tr~P}\exp\int^{\pi}_{-\pi}d\tau\left[\begin{array}{cc}
i\dot{X}^{\mu}A^{(1)}_{\mu}-2i\psi^{\mu}\psi^{\nu}F^{(1)}_{\mu\nu}~&~0
\nonumber\\
0~&~i\dot{X}^{\mu}
A^{(2)}_{\mu}-2i\psi^{\mu}\psi^{\nu}F^{(2)}_{\mu\nu}\nonumber
\end{array}\right]\\
&=&\int DX D\psi \exp[-I_0(X,\psi)]\nonumber\\
&\times&\sum_{k=1}^{2}{\rm Tr~P}\exp\int^{\pi}_{-\pi}d\tau
[i\dot{X}^{\mu}A^{(k)}_{\mu}-2i\psi^{\mu}\psi^{\nu}F^{(k)}_{\mu\nu}].
\end{eqnarray}

This defines the sum of the NSNS action for 
$A^{(1)}_{\mu}(X), A^{(2)}_{\mu}(X)$. 
In general form of $A_{\mu}(X)$ this integration is only perturbatively 
possible, while in the approximation that $F_{\mu\nu}$ and the metric 
$g_{\mu\nu}$
are constant the integration is exact \cite{FrTs,Ts1,AnTs1} 
and $Z$ becomes as follows:
\begin{eqnarray}
Z =T_9 \int d^{10}x \left[\sqrt{-\det\{g_{\mu\nu}+
4 \pi ( B_{\mu\nu} + F^{(1)}_{\mu\nu})  \}} 
+ \sqrt{-\det\{g_{\mu\nu}+
4 \pi ( B_{\mu\nu} + F^{(2)}_{\mu\nu}) \}}\right],
\end{eqnarray}
where $T_9$ is tension of a BPS D9-brane.
Here we have replaced $F^{(i)}$ 
with $F^{(i)}+ B$ 
using the $\Lambda$ symmetry, 
$F^{(i)}_{\mu \nu} \rightarrow F^{(i)}_{\mu \nu} +(d \Lambda)_{\mu \nu} $
and $B_{\mu \nu} \rightarrow B_{\mu \nu} -(d \Lambda)_{\mu \nu} $
for later convenience.
This is the desired result ($Dp-\overline{Dp}$ case is obtained
by T-duality).

\section{Exact Tachyon Condensation on Brane-Antibrane Systems}
\setcounter{equation}{0}

Here we investigate the tachyon condensation on brane-antibrane systems. 
It is obvious that the condensation of a constant tachyon field leads to the 
decay into the closed string vacuum $S=0$. The more interesting process is the
condensation of topologically nontrivial configurations of the tachyon
field. According to the Sen's conjecture \cite{sen14,Wi1} a codimension
$n$ configuration on D$p$-branes generally produces  D$(p-n)$-branes. As
we will see below, from the tractable free field calculations in BSFT we 
can describe such a tachyon condensation exactly. 
We can perform explicit computations 
in parallel with that for non-BPS D-branes discussed in 
\cite{KuMaMo2}. A crucial difference is that we can allow the
vortex-type configurations since the tachyon field on a brane-antibrane
system is a complex scalar field.

Note also that even though our regularization throughout in this paper
is based on ``$\ep$ -prescription" used in \cite{AnTs1,Ts1,Ts2,An}, 
the result does not change if we use
the point splitting regularization as in \cite{Wi2, KuMaMo2}.

First let us consider the condensation of the vortex-type tachyon field
on a single $\mbox{D}p-\overline{\mbox{D}p}$. From the viewpoint of the
boundary conformal field theory (BCFT) one can describe the
condensation as a marginal deformation \cite{MaSe2,NaTaUe} only. 
In BSFT defined by the boundary action (\ref{eq6}),
we can also handle a relevant perturbation
with two real parameters $u_i\ \ (i=1,2)$ :
\ba
T(X)=\f{1}{2}(iu_1X_1+u_2X_2).\label{tak}
\ea
Here we have set the gauge fields to zero. 
One can show that any tachyon fields of the form
$T(X)=a+\sum_{\mu=0}^{p}b_\mu X^\mu,\ \ (a,b_\mu \in {\mb{C}})$ can be
put into the form (\ref{tak}) 
by a Poincar\'e transformation and the U(1) gauge transformation.

Note also that this perturbation can be treated as a free boundary interaction 
and therefore the mixings with the other open string modes are avoided 
for all values of $u_i$ (see also \cite{KuMaMo1,KuMaMo2}).

For non-zero $u_i$ this represents a vortex which is a codimension two configuration along the $x^1$-$x^2$ plane. If $u_1=0$ or $u_2=0$, then this corresponds to a kink configuration. This fact is easy to see if one notes that the bosonic zero mode 
structure of BSFT on the $\mbox{D}p-\overline{\mbox{D}p}$ behaves as
\ba
S\sim e^{-|T(x_1,x_2)|^2}=e^{-\f{1}{4}(u_1x_1)^2-\f{1}{4}(u_2x_2)^2},
\ea
which shows the tachyon condensation leads to the desired localized
configuration for large $u_1$ and $u_2$.

In the presence of this boundary interaction the correlation functions
in the NS-sector are given by \cite{KuMaMo2}
\ba
&&G_B(\tau-\tau ',y_i,\ep)=<X^i(\tau)X^i(\tau ')>=2\sum_{m\in {\mb{Z}}}
\f{1}{|m|+y_i}e^{im(\tau-\tau ')-\ep|m|}, \label{gb}\\
&&G_F(\tau-\tau ',y_i,\ep)=<\psi^i(\tau) \psi^i(\tau ')>
=-\f{i}{2}\sum_{r\in
{\mb{Z}}+\f12}\f{r}{|r|+y_i}e^{ir(\tau-\tau')-\ep|r|}, \CR
&& \hspace{9cm} ( i \, : \, \mbox{no sum})
\label{gf}
\ea
where we have defined 
\ba
y_i=u_i^2.
\ea
Note that in the above expression we have used ``$\ep$ -regularization" discussed in \cite{AnTs1,Ts1,Ts2,An}.

%Note that here we have incorporated the bosonic zero modes as opposed to section%2.

On the other hand the BSFT action $S(=Z)$ including the boundary interaction can be computed by differentiating with respect to $y_1$ and $y_2$, respectively 
\ba
\f{\de}{\de y_i}\log S&=&-\f{1}{8\pi}\int_{0}^{2\pi} d\tau <X^i(\tau)X^i(\tau)
-4\psi^i(\tau)\f{1}{\de_\tau}\psi^i(\tau)>\no
&=&-\f{1}{2}[G_B(0,y_i,\ep)-G_B(0,2y_i,\ep/2)]. \label{des}
\ea

The difference of the correlators $G_B(0,y_i,\ep)-G_B(0,2y_i,\ep/2)$ is given in the limit of $\ep\to 0$ by
\ba
& &\lim_{\ep\to 0}[G_B(0,y,\ep)-G_B(0,2y_i,\ep/2)]=
4\sum_{m\ge 1} \left( \f{1}{m+y}
e^{-\ep m}-\f{1}{m+2 y} e^{-\f{\ep}{2}m} \right)+(\f{2}{y}-\f{1}{y})\no
&=&\lim_{\ep\to 0}[4\sum_{m\ge 1}\f{1}{m}(e^{-\ep m}-e^{-\f{\ep}{2}m})]+
4\sum_{m\ge 1}(\f{1}{m+y}-\f{1}{m})-4\sum_{m\ge 1}(\f{1}{m+2y}-\f{1}{m})+
(\f{2}{y}-\f{1}{y})\no
&=&-4\log 2-(4\f{d}{dy}\log\Gamma(y)+\f{2}{y}+4\gamma)+(2\f{d}{dy}\log
\Gamma(2y)+\f{1}{y}+4\gamma),
\ea
where we have used the following formulae:
\ba
& &\sum_{m\ge 1}\f{1}{m}e^{-\ep m}=-\log \ep+O(\ep), \\
& &\f{d}{dy}\log\Gamma(y)=-\f{1}{y}+\sum_{m\ge 1}\f{y}{m(m+y)}-\gamma\ \ \
(\mbox{$\gamma$: Euler's constant}).
\ea
Then it is easy to integrate eq.(\ref{des}) and we obtain $S$ up to the overall normalization $S_0$:
\ba
&& S(y_1,y,2)=S_0Z(y_1)Z(y_2), \no
&& Z(y)=4^{y}\f{Z_1(y)^2}{Z_1(2y)},
\ea
where $Z_1$ is a function peculiar to BSFT \cite{Wi2},
\ba
Z_{1}(y)=\s{y}e^{\gamma y}\Gamma (y).
\ea
The original $\mbox{D}p-\overline{\mbox{D}p}$ corresponds to $u_i=0$ and 
at this value the action is divergent since the world-volume of the brane 
is non-compact :
\ba
S(\mbox{D}p-\overline{\mbox{D}p})\to S_0\f{2}{\s{y_1y_2}}\ \ \ (y_i\to 0).
\ea
In the above computation the parameters $y_i$ play the role of cutoffs and are
 equivalent to a regularization by compactification $X^i\sim X^i+R_i$ as
\ba
\f{R_1R_2}{2\pi}\sim\int_{0}^{\infty}\f{dx^1dx^2}{2\pi}e^{-\f{1}{4}y_1(x^1)^2
-\f{1}{4}y_2(x^2)^2}=\f{2}{\s{y_1y_2}}.
\ea 
Let us now condense the tachyon field. 
Only when the tachyon is infinitely condensed $u_i=\infty$, 
the conformal invariance is restored,
which implies that the equation of motion is satisfied,
because $Z(y)$ is a monotonically decreasing function of $y$.
Therefore there exist three decay modes 
$(u_1,u_2)=(\infty,0),\ (0,\infty),\mbox{or}\ (\infty,\infty)$. 
The first two cases represent the kink configurations 
and we expect a non-BPS D$(p-1)$-brane will be generated 
at $x^1=0$ or $x^2=0$, respectively. This speculation is verified 
if one computes the tension (for $(u_1,u_2)=(\infty,0)$) 
and see that the correct value\footnote{Note that we set $\al=2$ 
and that the tension of a non-BPS Dp-brane is larger than 
that of a BPS D-brane by the factor $\s{2}$.} is reproduced as follows
\ba
\f{T_{Dp-\bar{D}p}}{T_{D(p-1)}}=\f{S(0,0)\cdot (R_1)^{-1}}{S(\infty,0)}=
\f{1}{2\pi},
\ea
where $T_{Dp-\bar{D}p}$ and $T_{D(p-1)}$ denotes
 the tension of a $\mbox{D}p-\overline{\mbox{D}p}$ 
and a non-BPS D$(p-1)$-brane, respectively; we have also used the fact
\ba
Z(y)\to \s{2\pi}\ \ \ (y\to\infty).
\ea

More intuitive way to see the generation of a non-BPS D$(p-1)$-brane is
to discuss the boundary interaction (\ref{eq6}). Let us shift the
original tachyon field by a real constant $T_0$ along $x^1$ as follows:
\ba
T(X)=\f{1}{2}T_0+\f{i}{2}u_1X_1. \label{tachyon} 
\ea
%and assume $A^{(-)}=0$ for simplicity. 
Then the boundary interaction (\ref{eq6}) after the 
condensation of the tachyon field (\ref{tachyon}) becomes
\ba
I_{B}&=&\int_{\de\Sigma} d\tau [\bar{\eta}\dot{\eta}+\f{1}{\s{2\pi}}u_1\psi^1 
(\eta-\bar{\eta})+\frac{1}{8\pi}T_0^2+\frac{1}{8\pi}u_1(X^1)^2+\ddd], \label{ib}
\ea
where the new tachyon field $T_0$ depends only 
on $X^a\ \ (a=0,2,\ddd,p)$. 
{}From this expression it is easy to see that 
in the limit of $u_1\to\infty$ 
we can set $\eta=\bar{\eta}$ after we perform 
the path integral of the fermion $\psi^1$. Then the term in $\ddd$ which depends on 
$A_{\mu\nu}^{(-)}$ vanishes because it is proportional to 
$\bar{\eta}\eta\sim 0$. On the other hand, the gauge field
$A_{\mu}^{(+)}$ is not sensitive to the tachyon condensation except
that the element $A_{1}^{(+)}$ is no longer a gauge field  but a transverse scalar field since the boundary condition along $x^1$ becomes Dirichlet. Thus the final boundary action after integrating out the fields $X^1,\psi^1$ is identified with that of a non-BPS D$(p-1)$-brane (\ref{bin}).

Next we turn to the last case $(u_1,u_2)=(\infty,\infty)$. This
corresponds to the vortex-type configuration and a BPS D$(p-2)$-brane is
expected to be generated at $(x^1,x^2)=(0,0)$. This fact is also checked
by comparing the tension 
as follows:
\ba
\f{T_{Dp-\bar{D}p}}{T_{D(p-2)}}=\f{S(0,0)\cdot (R_1R_2)^{-1}}{S(\infty,\infty)}=\f{1}{4\pi^2},
\ea
matching with the known result. Also note that this configuration has no tachyonic modes as desired. Indeed constant shifts of the original tachyon field 
(\ref{tak}) are equivalent to the shift of the position where the D-brane is generated.

It is also interesting to consider multiple branes and antibranes. This can be 
represented by the Chan-Paton factors. Following \cite{Wi1} let us consider 
$2^{k-1}$ pairs of brane-antibranes and condense the tachyon field
\ba
T(X)=i\f{u}{2}\sum_{\mu=1}^{2k}\Gamma^\mu X_\mu,\label{abs-1}
\ea
where $\Gamma^\mu$ denote $2^{k-1}\times 2^{k-1}$ $\Gamma$-matrices in $2k$ dimension and the extra factor $i$ is due to our convention of $\Gamma$-matrices.
%${\Gamma^\mu,\Gamma^\nu}=-2\eta^{\mu\nu}$

These configurations carry K-theory charges known as Atiyah-Bott-Shapiro 
construction \cite{ABS} and a BPS D$(p-2k)$-brane is expected to be generated.
 This fact will be more explicit by investigating the RR couplings in
the next section.
 The verification of the correct tension is the same as in the previous
cases if one notes that the additional factor $2^{k-1}$ from the
Chan-Paton factor 
should be included. 
Similarly one can also see the condensation of the tachyon field 
\ba
T(X)=i\f{u}{2}\sum_{\mu=1}^{2k-1}\Gamma^\mu X_\mu, 
\ea
on $2^{k-1}$ pairs of brane-antibranes produces a non-BPS D$(p-2k+1)$-brane.

In this way we have obtained all decay modes which can be handled in BSFT by free field calculations and these are all consistent with the BCFT results and 
 K-theoretic arguments. The incorporation of $B$-field (or equally $F^{(+)}$) can also be performed by free field calculations and in the same way as in 
 \cite{Ok,An,NeYa,Co1,LiWi} we have only to replace the parameters as follows 
\ba
y\equiv\mbox{diag}(y_1,y_2)\ \ \to \ \ \f{y}{1+2\pi B}.
\ea
This explains the extra factor $\s{\mbox{det}(1+2\pi\al B)}$ of 
the $\mbox{D}p-\overline{\mbox{D}p}$ tension in the presence of the $B$-field.

\section{RR Couplings and Superconnection}
\setcounter{equation}{0}

In this section we compute the RR couplings on non-BPS D$p$-branes and
brane-antibrane systems (D$p-\overline{\mbox{D}p}$) in a flat space
within the framework of BSFT. Since the backgrounds of the closed string should be on-shell in BSFT, we can only consider RR-fields $C_{\mu_1\ddd\mu_p}$ which obey the equation of motion. This is sufficient to determine the RR couplings of D-branes. 

Some of the RR couplings from the considerations of the descent relations 
\cite{sen16, sen18} and 
from the calculations of on-shell scattering amplitudes were already
obtained in the literature \cite{BiCrRo} for non-BPS D-branes and
\cite{KeWi} for brane-antibrane systems. However our off-shell
calculations 
in BSFT give a more powerful and unified viewpoint as we will see. For
example our method determines all the unknown coefficients of the higher
order terms with respect to $dT$ for non-BPS D-branes. Furthermore the
resulting expressions in both systems can be identified with an
intriguing mathematical structure known as
superconnection \cite{Qu}. This fact was already conjectured in
\cite{Wi1,KeWi} 
for brane-antibrane systems. Here we find the explicit proof of this in
BSFT and we point out that this structure can also be found in the RR
couplings on non-BPS D-branes. These results give another evidence of
the K-theory classification 
of D-brane charges \cite{MiMo,Wi1,Ho1}.

In the first two subsections we assume $p=9$. In the last subsection we determine the RR couplings for any $p$ including the effects of non-abelian transverse scalars. As a result we obtain the complete forms of Myers terms \cite{My} for both non-BPS D-branes and brane-antibrane systems. 

\subsection{RR Couplings on Non-BPS D-branes and Brane-Antibrane System in BCFT}
We regard the small shifts of the RR backgrounds as the perturbations. These shifts are realized in BSFT as the insertions of the RR vertex operators in the disk $\Sigma$. The picture \cite{FrMaSh} of the vertex operators\footnote{For 
 some subtleties, we recommend the readers to refer to \cite{BiDiFrLePeRuSc}.}
should be $(-\f{1}{2},-\f{3}{2})$. This is because here we consider 
only one insertion of them and because the total picture number 
on the disk should be $-2$.
 We mainly follow the conventions in \cite{GaMy}, where the scattering
 amplitudes of closed string from D-branes were computed. 
The RR vertex operators are given by
\ba
V^{(-\f{1}{2},-\f{3}{2})}&=&
e^{-\f12\phi-\f32\ti{\phi}}(P_{-}\hat{C})^{AB}S_A\ti{S}_B,  \\
\hat{C}^{AB}&=&\f{1}{p!}(\Gamma^{\mu_1\ddd\mu_p})^{AB}C_{\mu_1\ddd\mu_p}, 
\label{rr0}
\ea
where $\phi,\ti{\phi}$ denote the bosonized superconformal ghost of
 left-moving and right-moving sector, respectively; we define $S_A,\
 \ti{S}_B$ as the spin-fields of world-sheet fermions and $P_{-}$
 denotes the projection of the chirality. For more details of the
 notation for spinors see the appendix.

In the definition $S=Z$ of BSFT on non-BPS D-branes and brane-antibrane
systems the ghost sectors are neglected in the same way as in the case 
of the BPS D-branes discussed in \cite{Ts1,AnTs1}. 
Therefore it is difficult to
handle the ghost parts of the above RR vertex operators
explicitly. However it is natural to consider that the ghost parts and
the matter parts are decoupled and that the ghost parts give the trivial
contribution in the present calculations on the disk. Thus we can
compute RR couplings taking only the matter parts into
consideration.

Next we discuss the supersymmetry in the boundary interactions.
The supersymmetry is completely preserved in the one dimensional
boundary theory since all fermions at the boundary of the disk obey
periodic boundary conditions 
due to the cut generated by the RR vertex. Therefore one can
believe that the contributions from fermions and bosons are canceled
with each other for nonzero-modes and that the boundary theory becomes
topological in the sense of \cite{OoOzYi}. Note that in this paper we
consider only D-branes in a flat space and we have no corrections 
from world-sheet instantons.

First we turn to non-BPS D9-branes and determine the RR couplings up to the overall normalization. To see the bose-fermi cancellation explicitly let us assume that the tachyon field $T(X)$ is a linear function as $T(X)
=T_0+u_{\mu}X^{\mu}$ and the field strength $F_{\mu\nu}$ is constant. Then the
boundary interactions (\ref{bin}) are described as a free
theory. Furthermore in the R-sector the zero-modes and nonzero-modes are
completely decoupled and it is easy to see the bose-fermi cancellation
for nonzero-modes because of the supersymmetry as follows
\ba
\f{\de}{\de (u_\mu)^2}\log Z=-\f{1}{8\pi}\int_{0}^{2\pi} d\tau 
<\xi^\mu(\tau)\xi^\mu(\tau)
-4\psi^\mu(\tau)\f{1}{\de_\tau}\psi^\mu(\tau)>=0. \;\; (\mu \, : \mbox{no sum})
\ea
Note that this property in the R-sector is in strikingly contrast with the results (\ref{des}) in the NS-sector.

Thus we have only to discuss the bosonic and fermionic zero-modes. The path integral of the former is written as an integral over the world-volume coordinates $x^0,\ddd,x^9$. The latters are divided into that of the world-sheet fermions $\psi^\mu$ and of the boundary fermion $\eta$. The integral of the zero modes of $\psi^\mu$ in the action (\ref{bin}) can be replaced with the trace over $\Gamma$-matrices in Hamiltonian formalism as follows:
\ba
\psi^{\mu}\ \to\ \f{1}{\s{2}}i^{\f12}\Gamma^{\mu},
\label{co}
\ea
where the factor $i^{\f12}$ is due to the conformal map from the open string picture to the closed string picture\footnote{See for example \cite{CaLaNaYp}.}.
Furthermore, we can compute the contribution from the boundary fermion in
Hamiltonian formalism and its quantization is given by
$\hat{\eta}^2=\f{1}{4}$. Notice that we should assume $\Gamma^\mu$ and
$\hat{\eta}$ do anti-commute because in eq.(\ref{co}) we have not
included a cocycle factor. 
Then the result is given as follows
including the RR vertex operator :
\ba
S=\ti{\mu} \ \mbox{Tr}\int d^{10} x [\ :\exp[-\f14
T^2-2i^{\f32}\s{\pi}\Gamma^{\mu} \hat{\eta}\de_{\mu}T+2\pi
F_{\mu\nu}\Gamma^{\mu}\Gamma^{\nu}]:(P_{-}\hat{C})\ \hat{\eta}\ ] , 
\label{rr1}
\ea
where $\ti{\mu}$ represents the overall normalization and Tr denotes a trace with respect to both the $\Gamma$-matrices and the boundary fermion $\hat{\eta}$; the symbol $:\ :$ means that $\Gamma$-matrices are antisymmetrized because any operators should be normal-ordered in Hamiltonian formalisms. Note that an extra zero mode of $\eta$ is inserted due to its periodic boundary condition.

After we take the trace using the famous relation (\ref{c3})
between the Clifford
algebra (\ref{c2}) and the differential forms (\ref{c1}) and recover
$\al=2$, we easily obtain the final expression of the RR coupling on a
non-BPS D9-brane\footnote{The leading term $\sim \int C\we e^{-\f14
T^2}dT$ was already pointed out in \cite{KuMaMo2}.} as we will show in
the appendix. Its non-abelian generalization is also straightforward
using the expression (\ref{nb-n}) and one has only to add the trace of
the Chan-Paton factor in front of the above expression. Thus the result
is given up to the overall factor $\mu '$\footnote{Note that the above
action is real if only and only if $\mu '$ is proportional to
$i^{-\f32}$. Later we will determine this as $\mu'= - i^{-\frac{3}{2}}
T_{9}/\sqrt{2}$.}
by:
\ba
S&=&\mu '\ \mbox{Tr}_{\sigma}\ [\ \int C\wedge \exp{[-\f{1}{4}T^2-\s{\f{\pi\al}{2}}i^{\f32}DT\sigma_1+2\pi\al F]}\sigma_1], \no \label{wz-n}
 &=&i\mu '\ \mbox{Tr}\int C\wedge \exp{[-\f{1}{4}T^2-\s{\f{\pi\al}{2}}i^{\f12}DT+2\pi\al F]}\ |_{odd},
\ea
where the Pauli matrix $\sigma_1$ is equivalent to the boundary fermion
as $\sigma_1 \simeq 2\hat{\eta}$ and the trace $\mbox{Tr}_{\sigma}$ in
the first equation also
involves this freedom; the covariant derivative of the Hermitian tachyon
field on the non-BPS D-branes is denoted by $DT=dT-i[A,T]$. Also note
that in the second expression only the terms which include the odd
powers of $DT$ should be remained because of the trace with respect to
the boundary fermion and therefore we have represented this prescription
by $|_{odd}$ . 
{}From the above arguments,
we can see that the boundary fermion $\eta$ plays a crucial role in the
computations of the RR couplings.

Next let us discuss the RR couplings of a D$9-\overline{\mbox{D}9}$ in
BSFT. 
In this case decoupling of the zero-modes and nonzero-modes 
should also occur. 
Even though it is not so easy to give the explicit proof of the
bose-fermi cancellation in this case, it is natural to assume this
cancellation. The path integral of the boundary fermions
$\eta,\bar{\eta}$ can be represented 
by the RR-sector analog of the important formula (\ref{eq9}):
\ba
& &\int D\eta D\bar{\eta}\ e^{-I_{B}}=\mbox{Tr P}\ (-1)^F\ \exp\ \int_{-\pi}^{\pi}d\tau {\mb{M}}(\tau), \label{rr2} \\
& &{\mb{M}}(\tau)=
\left(
 	\begin{array}{cc}
 	i\dot{X}^\mu A^{(1)}_{\mu}-2i\psi^{\mu}\psi^{\nu}F^{(1)}_{\mu\nu}
 	-\f{1}{2\pi}T\bar{T} & i\s{\f{2}{\pi}}\psi^{\mu}D_{\mu}T \\
 	-i\s{\f{2}{\pi}}\psi^{\mu}\overline{D_{\mu}T} & i\dot{X}^\mu A^{(2)}_{\mu}
 	-2i\psi^{\mu}\psi^{\nu}F^{(2)}_{\mu\nu}-\f{1}{2\pi}\bar{T}T
 	\end{array}
 \right), \nonumber
\ea
where the insertion of $(-1)^F (=[\bar{\eta},\eta])$ is due to the
periodic boundary condition of $\eta,\bar{\eta}$ and can be replaced
with the Pauli matrix $\sigma_3$. Since we have only to take the zero
modes into account, we can regard the path-ordered
trace Tr P as the ordinary trace and thus we obtain
\ba
\int D\eta D\bar{\eta}DXD\psi\ e^{-I_{0}-I_{B}}=\mbox{Str}\ \exp
\left(
	\begin{array}{cc}
	2\pi F_{\mu\nu}^{(1)}\Gamma^{\mu}\Gamma^{\nu}-T\bar{T} & 2(i)^{\f32}\s{\pi}\Gamma^{\mu}D_{\mu}T \\
	-2(i)^{\f32}\s{\pi}\Gamma^{\mu}\overline{D_{\mu}T} & 2\pi F_{\mu\nu}^{(2)}\Gamma^{\mu}\Gamma^{\nu}-\bar{T}T
	\end{array}
\right),
\ea
where a supertrace Str is defined to be a trace with the insertion of $\sigma_3$. 

After we insert the RR vertex and again replace products of
$\Gamma$-matrices with differential forms, we obtain the following RR
couplings on a $\mbox{D}9-\overline{\mbox{D}9}$ :
\ba
S=\mu ''\ \mbox{Str}\ \int C\wedge \exp
\left(
	\begin{array}{cc}
	2\pi\al F^{(1)}-T\bar{T} & (i)^{\f32}\s{2\pi\al}\ DT \\
	-(i)^{\f32}\s{2\pi\al}\ \overline{DT} & 2\pi\al F^{(2)}-\bar{T}T
	\end{array}
\right), \label{wz}
\ea
where $\mu ''$ is a real constant\footnote{Even though one may
 think the factor $i^{\f32}$ strange at first sight, it is an easy task
to show that the action is indeed real by expanding the exponential.
Later we will determine this as $\mu''=T_9$}. This
result coincides with the proposal in \cite{KeWi} as we will 
see in the next subsection.

The non-abelian generalization is also straightforward if the above
abelian supertrace is replaced with the non-abelian one:
\ba
\mbox{Str diag}(a_1,a_2,\ddd,a_N,b_1,b_2,\ddd,b_M)=\sum_{i=1}^N a_i-
\sum_{j=1}^M b_j,
\ea
where we assume that there are $N$ D$9$-branes and $M$ antiD$9$-branes. This
result coincides with the proposal in \cite{KeWi} including the numerical
factors as we will see in the next subsection.

In this way we have derived the explicit RR couplings on a non-BPS
D9-brane and a $\mbox{D9}-\overline{\mbox{D9}}$ system in BSFT. The point is that one can
read off the RR couplings if one extracts the fermionic zero modes from
the boundary action $I_{B}$. This may be said as a boundary topological
model which can naturally lead to the notion of superconnection as we
will see in the next subsection. Note also that the above results can be
applied for general $p$-brane if the transverse scalars are set to zero.

\subsection{Superconnection and K-theory Charge}

Here we discuss the interpretation of the RR couplings on non-BPS
D-branes and brane-antibrane systems in BSFT as superconnections
\cite{Qu}. For brane-antibrane systems this fact was first suggested in
\cite{Wi1}. A definite relation between the RR couplings and the Chern
character of the superconnection was proposed in the paper
\cite{KeWi}. Our calculations in the previous subsection show that this
interpretation indeed holds within the framework of BSFT as we will see
below. Moreover we argue that such an interpretation can be applied to
non-BPS D-branes and our previous calculations give an evidence for
this.

Let us first review the definition and properties of superconnection following
 \cite{Qu}. There are two kinds of superconnections: one is for even-cohomology
 and the other is for odd-cohomology. In the K-theoretic language 
 the former is related to $K^{0}(M)$ and the latter to $K^{1}(M)$, where $M$ is  a manifold regarded as the D-brane world-volume. Both are defined 
 as follows\footnote{We include the explicit factor $\f{i}{2\pi}$ in front of 
 the field strength which was omitted for simplicity in the original paper \cite{Qu}. This is the reason why the factor $i^{\f12}$ does appear in the 
 expressions below.}:

\begin{flushleft}
\noindent{\bf Superconnection for $K^{0}(M)$}
\end{flushleft}
In this case we consider the $\zz$-graded vector bundle $E=E^{(0)}\oplus
E^{(1)}$, which can be directly applied to a brane-antibrane system if
one identifies $E^{(0)}$ and $E^{(1)}$ as the vector bundle on the branes
and antibranes, respectively. Then the endomorphism of this superbundle
$X\in \mbox{End}\ E$ has the following $\zz$-grading:
\ba
\mbox{deg}(X)=\left\{\begin{array}{c}
\ 0 \ \ \mbox{if}\ X:E^{(0)}\to E^{(0)}\ \mbox{or}\ E^{(1)}\to E^{(1)},  \\
\ 1 \ \ \mbox{if}\ X:E^{(0)}\to E^{(1)}\ \mbox{or}\ E^{(1)}\to E^{(0)}.
\end{array} \right .
\ea

In addition, there is also a natural $\z$-grading $p$ if one considers
the algebra of the differential forms $\Omega(M)=\oplus\Omega^{p}(M)$,
where $\Omega^{p}(M)$ denotes the algebra of $p$-forms on $M$. The
crucial observation is to mix these two gradings and to define the
$\zz$-grading for 
$\alpha\in \Omega^{p}(M,\mbox{End}\ E)=
\Omega^{p}(M)\otimes\Omega^{0}(M,\mbox{End}\ E)$ as follows: 
\ba
\alpha=\omega\otimes X\in\Omega^{p}(M)
\otimes\Omega^{0}(M,\mbox{End}\ E),\ \ 
\mbox{deg}(\alpha)\equiv p+\mbox{deg}(X), \label{super1}
\ea
where $\Omega^{0}(M,\mbox{End}\ E)$ denotes the space of sections of $\mbox{End}\ E$.
Then the superalgebra is defined by the following rule:
\ba
(\omega\otimes X)(\eta\otimes Y)=(-1)^{\mbox{deg}(X)\cdot\mbox{deg}(\eta)}\ 
(\omega\eta\otimes XY), \label{super2}
\ea
and the supercommutator can be defined as:
\ba
[\ap,\beta]=\ap\beta-(-1)^{\mbox{deg}(\alpha)\cdot\mbox{deg}(\beta)}\beta\alpha.\label{super3}
\ea

An element of $\Omega^{p}(M,\mbox{End}\ E)$ can be written as a $2\times
2$ matrix, where the diagonal elements and off-diagonal elements have
even and odd degree of $\Omega^{0}(M,\mbox{End}\ E)$, respectively. We also define the supertrace as
\ba
\ap\in\Omega(M,\mbox{End}\ E)=
\left(
	\begin{array}{cc}
	\ap_1 & \ap_2 \\
	\ap_3 & \ap_4
	\end{array}
\right), \ \ \ \mbox{Str}(\ap)=\mbox{Tr}(\ap_1)-\mbox{Tr}(\ap_4)\in\Omega(M) ,
\ea
where $\mbox{Tr}$ denotes the ordinary trace of vector bundles. Note that the supertrace vanishes on supercommutators.

 Let us now define a superconnection on $E$ to be an operator 
${\ca{D}}=d+{\ca{A}}$ on $\Omega(M,E)$ of odd degree satisfying the
derivation property:
\ba
{\ca{D}}(\omega\phi)=(d\omega)\phi+(-1)^{deg(\omega)}\omega 
({\ca{D}}\phi),\ \ \ \ \omega\in\Omega(M),\ \ \phi\in \Omega(M,E).
\ea

For local calculations familiar for physicists one can regard ${\ca{A}}$ as a degree odd element of $\Omega(M,\mbox{End}\ E)$:
\ba
{\ca{D}}=d+{\ca{A}}=
\left(
	\begin{array}{cc}
	d+A^{(1)} & \s{2\pi i}T \\
	\s{2\pi i}\bar{T} & d+A^{(2)}
	\end{array}
\right), 
\ea
where the factor $\s{2\pi i}$ has been included for later convenience. The
diagonal parts $d+A^{(1)},d+A^{(2)}$ denote the ordinary gauge
connections of vector bundles $E^{(1)},E^{(2)}$, respectively. 
$T$ denotes a odd degree
 endomorphism of $E$. Notice that in this definition the exterior
derivative $d$ does {\it anti-commute} with any odd element in
$\mbox{End}\ E$.

Then the 
curvature ${\ca{F}}$ of a superconnection ${\ca{D}}$ is defined to be an even degree element of $\Omega(M,\mbox{End}\ E)$:
\ba
{\ca{F}}&=&{\ca{D}}^2=d{\ca{A}}+{\ca{A}}^2 \no
&=&
\left(
	\begin{array}{cc}
	F^{(1)}+2\pi iT\bar{T} & \s{2\pi i}DT \\
	\s{2\pi i}D\bar{T} & F^{(2)}+2\pi i\bar{T}T
	\end{array}
\right), \label{su-k0}
\ea
where we have defined $DT=dT+A^{(1)}T+TA^{(2)}$\footnote{Note that since $T$ is an odd element, it anti-commutes with any one form. Therefore we can say that $T$ does couple to the relative gauge field $A^{(1)}-A^{(2)}$.}.

The ``Chern character'' of this superconnection is given by
\ba
\mbox{Str}\exp(\f{i}{2\pi}{\ca{D}}^2)=\mbox{Str}\exp\f{i}{2\pi}{\ca{F}}.\label{ch-0}
\ea
It is easy to see that this is closed because
\ba
d\ (\mbox{Str}{\ca{D}}^{2n})=\mbox{Str}[{\ca{D}},{\ca{D}}^{2n}]=0.
\ea
Furthermore as shown in the main theorem in \cite{Qu}, its cohomology class does not depend on the choice of $T$. In other words, this Chern character defines the same element of $K^0(M)$ irrespective of $T$:
\ba
\mbox{Str}\exp(\f{i}{2\pi}{\ca{D}}^2)\simeq \mbox{ch}(E_1)-\mbox{ch}(E_2)\in H^{even}(M,{\mb{Q}})\cong K^0(M),\label{kk-0}
\ea
where $\mbox{ch}(E)$ denotes the ordinary Chern character and we have applied 
the Chern isomorphism, which states that the even cohomology and the 
K-group $K^0(M)$ are equivalent if ${\mb{Q}}$ is tensored.

\begin{flushleft}
\noindent{\bf Superconnection for $K^{1}(M)$}
\end{flushleft}
The first step to define the second superconnection is to regard a bundle $E$ as
 a module over the Clifford algebra $C_1={\mb{C}}\oplus {\mb{C}}\sigma_1$.
 In other words, we define the endomorphism of this superbundle as
 $\mbox{End}_{\sigma}\ E=\mbox{End}\ E\otimes C_1$. Let us call all
 elements which include $\sigma_1$ degree odd and the others degree
 even. In the physical context these correspond to the fields on non-BPS
 D-branes which belong to GSO odd and even sectors, respectively. The 
supertrace on $\mbox{End}_{\sigma} E$ is defined as follows:
\ba
\mbox{Tr}_{\sigma}(X+Y\sigma_1)=2 \mbox{Tr}(Y),\label{trs1}
\ea
where $X,Y\in \mbox{End}_{\sigma}\ E$ are degree even elements. 
Further we mix the degree of differential forms in the same way as in 
the previous case eq.(\ref{super1}),(\ref{super2}) and (\ref{super3}).

A superconnection on $E$ is defined locally to be an odd element as follows
\ba
{\ca{D}}=d+{\ca{A}}=d+A-\s{\f{i\pi}{2}}T\sigma_1,
\ea
where $A$ is an ordinary connection and $T$ is a self-adjoint 
endomorphism.
The curvature of this is also defined as
\ba
{\ca{F}}&=&{\ca{D}}^2=d{\ca{A}}+{\ca{A}}^2 \no
        &=&F-\s{\f{i\pi}{2}}DT\sigma_1-\f{i\pi}{2}T^2, \label{su-k1}
\ea
where we have defined $DT=dT-i[A,T]$.
Then the ``odd Chern character'' is given by
\ba
\mbox{Tr}_{\sigma}\exp(\f{i}{2\pi}{\ca{D}}^2)=\mbox{Tr}_{\sigma}\exp\f{i}{2\pi}{\ca{F}}.\label{ch-1}
\ea
The main theorem in \cite{Qu} again tells us that this character is closed and its cohomology class does not depend on the choice of $T$. Further we can regard this as an element of K-theory group $K^1(M)$:
\ba
\mbox{Tr}_{\sigma}\exp(\f{i}{2\pi}{\ca{D}}^2)\ \in\ H^{odd}(M,{\mb{Q}})\cong K^1(M).
\ea

\begin{flushleft}
\noindent{\bf Physical interpretations}
\end{flushleft}

It is a well-known fact that the RR couplings on a BPS D9-brane 
are written by using Chern characters \cite{Li,Do1,GrHaMo}
\ba
S&=& T_9 \ \mbox{Tr}\ \int_{M} C\wedge \exp{2\pi\al F}, \label{ch}
\ea
where we have assumed that the world-volume (=spacetime) $M$ is flat. 
As can be deduced from this \cite{MiMo}, the D-brane charges in type IIB 
were proposed to be regarded as an
element of K-theory group $K^{0}(M)$, which
is equivalent to the Chern character up to torsion via the Chern
isomorphism. This proposal was strongly convinced in the study of
tachyon condensation on brane-antibrane systems \cite{Wi1}. The original
definition of $K^{0}(M)$ is given by considering the equivalence class
of a pair of vector bundles $(E_1,E_2)$. This definition can be
naturally seen as a mathematical description of brane-antibrane
systems. Moreover it was pointed out that the other K-theory group
$K^{1}(M)$ is related to the tachyon condensation on non-BPS D9-branes
\cite{Ho1}. 
This leads to the classification of the D-brane charges in type IIA.
At first sight, there are two different physical observations about the
generation of K-theory charges: the K-theory charges from RR couplings on
a BPS D-brane and those from the tachyon condensation. Then it is
natural to ask if we can directly fill this gap in string field
theories. The answers to this question is yes in BSFT and the key is superconnections as we see below.

The role of superconnections in the D-brane physics is explicit if one
notes that the RR couplings on D9-$\overline{\mbox{D9}}$ systems and non-BPS
D9-branes can be expressed as wedge products of RR-fields and the Chern
characters of superconnections:

\ba
S&=&\mu ''\ \mbox{Str}\ \int_{M}\ C\wedge \exp({\f{i}{2\pi}\ca{F}})\ \ \
\mbox{(for a D9-$\overline{\mbox{D9}}$ )},
\label{surr-1} 
\\
S&=&\mu '\ \mbox{Tr}_{\sigma}\ \int_{M}\ C\wedge
\exp({\f{i}{2\pi}\ca{F}})\ \ \ \mbox{(for a non-BPS
D9-brane)},\label{surr-2}
\ea
where the curvature ${\ca{F}}$ in the first equation represents the
superconnection for $K^{0}(M)$ and in the second for $K^{1}(M)$. One of
the expression (\ref{surr-1}) was already proposed in
\cite{KeWi}. 
One can indeed transform these mathematical expressions
eq.(\ref{su-k0}),(\ref{su-k1}) into the physical ones
eq.(\ref{wz}),(\ref{wz-n}) by following the prescription\footnote{Also
note that for a brane-antibrane we need an additional minus sign in
front of $\overline{DT}$. This occurs due to the following reason. The
mathematical definition of the superconnection for $K^{0}(M)$ assumes
that an odd form anti-commutes with an  odd degree endomorphism as in eq.(\ref{super2}). On the other hand, in the physical expression eq.(\ref{wz}) it does commute.}:
\ba
D=d+A &\to& 2\pi\s{\al}D=2\pi\s{\al}(d-iA).
\ea
Note that if one assumes the descent relation \cite{sen16,sen18}, one can 
formally obtain the coupling (\ref{surr-2}) from (\ref{surr-1}). Here
the descent relation argues that one can reduce the degree of freedom on
a brane-antibrane to that on a non-BPS D-brane if one projects 
the Chan-Paton factor $\Lambda$ on a brane-antibrane by the following action:
\ba
(-1)^{F_{L}}\ :\ \Lambda\ \to\ \sigma_1\Lambda\sigma_1,
\ea 
where $F_{L}$ denotes the spacetime fermion number in the left-moving sector.
Thus we have proved that the proposal in \cite{KeWi} is correct if we
 consider brane-antibranes in BSFT. Also the second new expression
 (\ref{surr-2}) is interesting because this explicitly shows that we can 
obtain the odd forms which correspond to $K^{1}(M)$ by including the
Hermitian tachyon field. 
 
Then let us turn to the first question. In the expression (\ref{surr-1}) 
we can smoothly connect the following two regions 
through the process of a tachyon condensation. 
Before the condensation the
RR charge in (\ref{surr-1}) comes only from the gauge
field-strengthes. On the other hand when the tachyon maximally
condenses, the contribution from the tachyon field dominates. 
For a trivial example, in \cite{Qu} it was shown 
that if the tachyon field $T$ is
invertible at some regions in $M$ then the Chern character (\ref{ch-0})
does locally vanish there. 
Physically this is natural since the condensation of
a constant tachyon leads to the decay into the vacuum and the lower 
dimensional charges are generated only at the regions where $T$ is not 
invertible. The same theorem also holds for the odd case (\ref{ch-1}). 

A nontrivial example of the tachyon field which is not invertible is
 given by the Atiyah-Bott-Shapiro construction (\ref{abs-1}). The
 important point is that RR charges or equally K-theory charges do not
 change globally during the tachyon condensation as is shown in
 eq.(\ref{kk-0}) and thus the charges are quantized in off-shell regions.
 To give a more concrete picture, let us remind the calculation of the
 tachyon condensation in the previous section. If one calculates the RR
 coupling (\ref{surr-1}) for the vortex-type tachyon configuration
 (\ref{tak}) and integrates it over the world-volume, then it is easy to
 see that the RR charges are independent of the parameters $u_1,u_2$
 except the ``singular points" $u_1=0$ or $u_2=0$. 
This analysis of the tachyon condensation 
determines the values of $\mu'$ and $\mu''$ as 
$\mu'=-i^{-\frac{3}{2}} \frac{T_9}{\sqrt{2}}$ and $\mu''=T_9$.
Notice that the
 topology of the tachyon field $T$ becomes trivial at the points $u_1=0$
 or $u_2=0$ and therefore one can not regard $T$ as an element in the
 desirable endomorphism. 
In this way we can relate D-brane charges in
 the RR-couplings to D-brane charges due to tachyon condensations
 directly in BSFT.

\subsection{RR Couplings of Transverse Scalars: Myers Terms}

As we have seen above, the BSFT calculations determine the RR couplings
on the various systems of 9-branes exactly. If one wants to obtain the RR
couplings for  a single $p(\neq 9)$-brane, one has only to interpret the
RR-fields as their pull-backs (see eq.(\ref{pull})) to the D-brane
world-volume.

However their non-abelian generalizations (\ref{ch}),(\ref{surr-1}) and
(\ref{surr-2}) are incomplete for $p(\neq 9)$-branes from the viewpoint
of T-duality. As pointed out in \cite{My} for BPS Dp-branes, in
order to recover the T-duality symmetry we must take the transverse
scalars $\Phi^i\ \ (i=p+1,\ddd,9)$ into account. For example if a pair
of scalars is noncommutative $[\Phi^i,\Phi^j]\neq 0$, then D($p+2$)-brane charges emerge and therefore we should include this
effect. Those terms which represent such an effect are called Myers
terms and their structures were investigated in \cite{My,HaMi,JaMe,MuSu,GaMy2}. 
Here we argue that if one would like to determine all of the Myers terms for any D-branes, then one has only to compute RR couplings in BSFT including the transverse scalars. Here we set the value of B-field to zero.

First let us determine the Myers terms for non-BPS D-branes. 
We use the non-abelian boundary action (\ref{nb-n}) with an additional
term due to the transverse scalars
\ba
-i\int_{\de\Sigma}d\tau d\theta\ \bar{\mb{\Delta}}\Phi_i({\mb{X}}) D_{n}{\mb{X}}^i{\mb{\Delta}},
\ea
where we have defined the T-dualized covariant derivative as
$D_{n}=\de/\de\theta+i\theta\de/\de\sigma$; $\sigma$ is the world-sheet
coordinate transverse to the boundary $\de\Sigma$. Note that if one
wants to discuss BPS D-branes, one has only to set $T=0,\ \Gamma=0$ in
eq.(\ref{nb-n}). After we integrate out the 
auxiliary fields and extract the zero modes as was done in (\ref{rr1}), we obtain the additional terms:
\ba
:\exp[\ddd+2i^{\f12}\s{\pi}[\Phi_i,T]\Gamma^i\hat{\eta}-2i\pi
[\Phi_i,\Phi_j]\Gamma^i\Gamma^j-4\pi
D_{\mu}\Phi_i\Gamma^{\mu}\Gamma^i]:\ ,
\ea
where $\ddd$ denotes the contribution from (\ref{rr1}). When one
estimates the trace of $\Gamma$-matrices and is reminded of the calculations in the appendix, note that the matrices $\Gamma^i\ \ (i=p+1,\ddd,9)$ contract the indices of RR-fields $C_{\mu_1,\ddd,\mu_q}$ in contrast with the matrices $\Gamma^{\mu}\ \ (\mu=0,\ddd,p)$. Then we obtain the following additional RR couplings (Myers terms) to eq.(\ref{wz-n}):
\ba
S=\mu'\ \mbox{Tr}\int[\exp[\ \ddd+\s{\f{\pi\al}{2}}i^{\f12}[\io_{\Phi},T]\sigma_1-2\pi\al i\ \io_{\Phi}\io_{\Phi}-2\pi\al\io_{D\Phi}]\ \sigma_1 \wedge C],
\ea
where $\mbox{Tr}$ denotes both the trace with respect to Chan-Paton factors and the trace defined by eq.(\ref{trs1}); $\io_{\Phi}$ and $\io_{D\Phi}$ denote the interior product by ${\Phi}$ and ${D\Phi}$:
\ba
&&A=\f{1}{r!}A_{\nu_1,\nu_2,\dd,\nu_r}dx^{\nu_1}dx^{\nu_2}\dd dx^{\nu_r},\\
&\to& \io_{\Phi}A=\f{1}{(r-1)!}\Phi^iA_{i,\nu_2,\ddd,\nu_r}dx^{\nu_2}\dd dx^{\nu_p},\ \ \io_{D\Phi}A=\f{1}{(r-1)!}D_\mu\Phi^iA_{i,\nu_2,\ddd,\nu_r}dx^{\mu}dx^{\nu_2}\dd dx^{\nu_r}.\nonumber
\ea

In the above RR coupling the first term is peculiar to non-BPS D-branes. The second corresponds to the generation of higher dimensional D-brane charges due to the noncommutative transverse scalars. The last term changes the RR fields $C$ into their covariantized expression $\ti{P}[C]$ of the pull-backs $P[C]$ :
\ba
P[C]_{\mu_1,\ddd,\mu_q}=C_{\nu_1,\ddd,\nu_p}(\f{\de y^{\nu_1}}{\de x_{\mu_1}})\ddd
(\f{\de y^{\nu_q}}{\de x_{\mu_q}}),\label{pull}
\ea
where $y^{\mu}=x^{\mu}\ \ (\mu=0,\ddd,p)$ denote the coordinates of the
p-brane world-volume and we also define $y^{i}=-2\pi\al\Phi^i \ \
(i=p+1,\ddd,9)$. In addition ``covariantized" means that all derivatives
$\de_{\mu} \Phi^i$ in the above definition (\ref{pull}) should be
replaced with covariant derivatives 
$D_{\mu} \Phi^i$. Then the total RR-couplings are given by
\ba
S=i\mu'\ \mbox{Tr}[\int \ti{P}[e^{-2\pi\al i\ \io_{\Phi}\io_{\Phi}+
\s{\f{\pi \al}{2i}}[\io_{\Phi},T]}\wedge C]
\wedge e^{-\f14T^2-\s{\f{\pi\al i}{2}}DT+2\pi\al F}]\ |_{odd},
\ea
where the trace Tr is a symmetric trace with respect to
$[\Phi^i,\Phi^j],\ [\Phi^i,T],\ T^2,\ DT,\ F$ and $D\Phi^i$. For example
the term proportional to $[\io_{\Phi},T]$ was already pointed out in
\cite{JaMe}. On the other hand if we set $T$ to zero and neglect the
restriction to odd forms, then one gets the RR couplings for BPS
D$p$-branes, matching with the results in \cite{My}. In this way we have
determined the complete form of Myers terms for non-BPS D-branes in BSFT
and these include new terms which are higher powers of
$[\io_{\Phi},T]$. 
Note also that our calculations explicitly preserve
the T-duality symmetry $A_{\mu}\lr -\Phi_{i}$.

Then let us turn to the final task in this section: Myers terms in
brane-antibrane systems. In the same way as before we have only to add
the extra terms which involve transverse scalars. As a result the matrix
in the exponential of eq.(\ref{wz}) includes Myers terms as
follows:
\ba
\left(
	\begin{array}{cc}
	2\pi\al(F^{(1)}-i\io_{\Phi^{(1)}}\io_{\Phi^{(1)}}-\io_{D\Phi^{(1)}})
	-T\bar{T} 
	& (i)^{\f32}\s{2\pi\al}\{DT+i(\io_{\Phi^{(1)}T}-\io_{T\Phi^{(2)}})\}  \\
	-(i)^{\f32}\s{2\pi\al} \{\overline{DT}-i(\io_{\bar{T}\Phi^{(1)}}-\io_{\Phi^{(2)}\bar{T}})\}
	 & 2\pi\al(F^{(2)}-i\io_{\Phi^{(2)}}\io_{\Phi^{(2)}}-\io_{D\Phi^{(2)}})
	-\bar{T}T 
	\end{array}
\right),
\ea
where we interpreted $\mbox{Str}$ in eq.(\ref{wz}) as both the symmetric
trace with respect to Chan-Paton factors and the original
supertrace. Note that if one requires that the branes and the antibranes
always have the common world-volume, then we get
$\Phi^{(1)}=\Phi^{(2)}(=\Phi)$. In this case we can find intriguing
terms in the RR couplings:
\ba
S\sim \int_{\mbox{p-brane}}
C^{(p+3)}_{i,j,\ddd}\mbox{Tr}[\Phi^{i},T][\Phi^{j},\bar{T}] e^{-T\bar{T}}.
\ea

%%%%%%%%%%%%%%%%%%%%%%%%%%%%%%%%%%%%%%%%%%%%%%%%%%%%
\section{The World-Volume Action for Brane-Antibrane System}
%%%%%%%%%%%%%%%%%%%%%%%%%%%%%%%%%%%%%%%%%%%%%%%%%%%%

\setcounter{equation}{0}

In this section we compute the BSFT action for tachyons and gauge
fields in the brane-antibrane system perturbatively.

First, we will show below that
the action computed from (\ref{eq8}) 
can be written as
\beqa
Z &=& T_9 \int d^{10}x e^{- T \bar{T}(x)} 
\left[\sqrt{-\det\{g_{\mu\nu}
+4\pi F^{(1)}_{\mu\nu} (x) \}}+ \sqrt{-\det\{g_{\mu\nu}
+4\pi F^{(2)}_{\mu\nu} (x) \}} \right. \nonumber \\
&& \hspace{8cm}
\left. + \sum_{n=1}^\infty 
{\cal{G}}_{2n} ( F^{(i)}_{\mu \nu}, T,\bar{T} ,D_{\rho})
\right],
\label{Z1}
\eeqa
where ${\cal{G}}_{2n}$ is a $2n$-derivative term constructed from
$F^{(i)}_{\mu \nu}, T, \bar{T}$ and $2n$ covariant derivatives $D_{\mu}$.
Note that, for example, 
$F_{\mu \nu}^{(-)} T$ is regarded as a 2-derivative term
and should be included in ${\cal{G}}_{2}$
because of the identity $ [D_\mu, D_\nu] T = -i F_{\mu \nu}^{(-)} T$.
These ambiguities are reminiscent of
the case of the non-Abelian Born-Infeld action \cite{Ts4}
in which 
$[F_{\mu \nu} ,F_{\gamma \sigma}]=
i [D_\mu, D_\nu] F_{\gamma \sigma}$ was regarded as a derivative term.

What does this action represent? We first answer the question.
As shown in \cite{FrTs}, this action (\ref{Z1}) is regarded as 
an on-shell effective action for $T$ and $A_{\mu}$, in which 
the massive modes are integrated out, 
or an off-shell BSFT action with other modes than $T$ and $A_{\mu}$
setting to zero. However, if the one-point function of the massive
fields vanishes in all the off-shell region, then we can regard this
action as the off-shell BSFT action, in which the massive fields are
integrated out.

The explanation is as follows. First we expand the full string field
action by the power series of the massive fields ($\lambda^i$):
\ba
S[T,A_{\mu},\lambda^i]=S^{(0)}[T,A_{\mu}]+\lambda^i S^{(1)}_i[T,A_{\mu}]
+\lambda^i\lambda^j S^{(2)}_{ij}[T,A_{\mu}]+\cdots\cdots.
\ea
If we want to obtain the effective string field action including only
the tachyon and the gauge fields, then we integrate out the massive
fields. Note that ``integrate out'' means that we only insert the
solution of the equation of motion for $\lambda^i$ because we want to obtain 
{\sl the tree level} effective string field action. The general solution 
of the equation of motion for $\lambda^i$ is very complicated. However, if 
$S^{(1)}_i[T,A_{\mu}]$ (the one-point function of $\lambda^i$) vanishes, 
then we can easily find one solution of the equation of motion, that is 
$\lambda^i=0$. Therefore in this case we can regard the
$S^{(0)}[T,A_{\mu}]$ as the effective action, in which the massive
fields are integrated out. $S^{(0)}[T,A_{\mu}]$ is just the action we
are able to calculate from the renormalizable sigma model action where
the massive fields are set to zero. Therefore the crucial point to
obtain the effective action is whether $S^{(1)}_i[T,A_{\mu}]$ does
vanish or not. {}From the argument of \cite{KuMaMo1, KuMaMo2}, 
$S^{(1)}_i[T,A_{\mu}]$ vanishes at least at the conformal fixed point of 
the renormalization group flow (on shell point), however in the off
shell region it is non-trivial. In \cite{KuMaMo1} they say that it is
correct for the free sigma model action because from the relation of
BSFT:
\ba
\frac{\partial S}{\partial \lambda^i}=\beta^j G_{ij}(\lambda),
\ea
(where $G_{ij}(\lambda)$ is some positive definite metric) 
$\frac{\partial S}{\partial \lambda^i}|_{\lambda=0}$ vanishes. However,
in this argument, the only ambiguity is whether the non-diagonal elements
of $G_{ij}(\lambda)$, where $j$ corresponds to $T,A_{\mu}$ and i to the
massive fields, does vanish or not (note that $\beta^i|_{\lambda=0}=0$
($i$ : the massive fields) and $\beta^{T,A}|_{\lambda=0}\neq
0$.). However, from the success and correctness of the tachyon
condensation in \cite{KuMaMo1, KuMaMo2} we can consider that above facts 
hold in the case of the free sigma model action. Therefore if the above
facts hold not only for the free sigma model but also for our renormalizable
one (eq.(\ref{eq3})), $S^{(1)}[T,A_{\mu}]$ vanishes in off-shell regions and we can
regard eq.(\ref{Z1}) as the effective action, in which the massive
modes are integrated out. However, we can not assert that it is true. 
  
Now we prove eq.(\ref{Z1}).
We use
the matrix form of the world-sheet action
\beq
Z=\int DX D \psi {\rm exp} [ -I_0(X,\psi) ]
{\rm Tr \, P \, exp} \int^{\pi}_{-\pi}d\tau \left[\begin{array}{c c }
M_{11} & M_{12}\nonumber\\
M_{21} & M_{22} \nonumber\\
\end{array}\right],
\eeq
where
\beqa
\left[\begin{array}{c c }
M_{11} & M_{12}\nonumber\\
M_{21} & M_{22} \nonumber\\
\end{array}\right]
=
\left[\begin{array}{c c }
i\dot{X}^{\mu}A^{(1)}_{\mu}-2i\psi^{\mu}\psi^{\nu}F^{(1)}_{\mu\nu}
-\frac{ 1 }{2 \pi} T \bar{T}
& ~ i\sqrt{\frac{2}{\pi}}\psi^{\mu}D_{\mu}T
\nonumber\\
-i\sqrt{\frac{2}{\pi}}\psi^{\mu}\overline{D_{\mu}T}~ 
& i\dot{X}^{\mu}
A^{(2)}_{\mu}-2i\psi^{\mu}\psi^{\nu}F^{(2)}_{\mu\nu}
-\frac{ 1 }{2 \pi} \bar{T} T
\nonumber
\end{array}\right]. \\
\eeqa
Now we expand $X^{\mu}$ from zero mode $x^{\mu}$ 
as $X^{\mu}=x^{\mu}+\xi^{\mu}$
and path integrate $\xi^{\mu}$ and $\psi^{\mu}$
($\psi$ does not include the zero mode in this case.).
Then from the expansion
%we can expand $T\bar{T} (X) $ as 
$T \bar{T}(X) =
T \bar{T}(x) + \xi^{\mu} (\pa_\mu (T \bar{T}) )(x)+\cdots=
T \bar{T}(x) 
+ \xi^{\mu} (D_\mu (T \bar{T}) )(x)+\cdots$,
we can replace $T\bar{T} (X) $ in $Z$ by $T \bar{T}(x)$
if we remove
the derivative terms which can be included in ${\cal{G}}_{2n}$.
The $F_{\mu \nu}^{(i)} (X)$ terms can also be
replaced by $F_{\mu \nu}^{(i)} (x)$. 
Furthermore we see that 
the contributions from the off diagonal part of the matrix $M_{ab}$ 
have the form of ${\cal{G}}_{2n}$
since they can be expanded as 
$D_{\mu} T(X) =
D_{\mu} T(x) + \xi^{\nu} (\pa_\nu (D_{\mu} T) )(x)+\cdots$.
The terms $i \dot{X}^{\mu} A_{\mu}^{(i)}(X) $ 
in the diagonal part 
are combined with 
the other non-gauge covariant terms
to give gauge covariant ones.
Therefore we conclude that the action for
the brane-antibrane system becomes (\ref{Z1}).
Similarly, we can show that the action for a system 
of $n$ branes and $m$ anti-branes
becomes
\beqa
Z &=& T_9 \int d^{10}x  \left(  {\rm SymTr} \left[
e^{- T \bar{T}  }
\sqrt{-\det\{g_{\mu\nu}
+4\pi F^{(1)}_{\mu\nu}  \}} 
+ e^{- \bar{T} T  }
\sqrt{-\det\{g_{\mu\nu}
+4\pi F^{(2)}_{\mu\nu}  \} }  \right] \right. \nonumber \\
&& \hspace{7cm}
\left. + \sum_{n=1}^\infty {\cal{G}}_{2n} ( F^{(i)}, T,\bar{T} ,D_{\mu})
\right),
\label{Z2}
\eeqa
where ${\rm SymTr}$ denotes the symmetrized trace 
for $T \bar{T}, \bar{T} T$ and $F^{(i)}_{\mu \nu}$ \cite{Ts4}.

Now we will use this action in order to 
investigate the non-commutative soliton in the 
brane-antibrane system \cite{HaKrLaMa1,witten2,HaKrLa1}.
Here we simply assume that $S^{(1)}_{i}[T,A_{\mu}]$ is zero or derivative 
terms, then the action (\ref{Z2}) is exact even for the off shell fields.
According to the argument in \cite{SeWi}
if we include the background constant
$B$-field, then the propagator is modified 
and the action can be written as the non-commutative field theory
when we use 
the point splitting regularization for the world-sheet theory.
This will be true at least for the on-shell fields.
We assume here that this is also true for the off-shell action
since the evidences for this have been obtained
\cite{Co1,Ok,An}.
Then the non-commutative action for the brane-antibrane system 
with background constant $B$-field becomes 
the same form as (\ref{Z2}) where the product is 
$*$-product and closed string metric $g_{\mu \nu}$ 
and coupling $g_s$ 
are replaced by open string metric $G_{\mu \nu}$ and $G_s$, respectively.
We should also replace the field strength $F^{(i)}_{\mu \nu}$ 
by $\hat{F}^{(i)}_{\mu \nu}+\Phi_{\mu \nu}$,
where $\p_{\mu \nu}$ represents a 
freedom to relate closed string quantities to open string quantities. 
Below we take $\Phi_{\mu \nu}=-B_{\mu \nu}$ for simplicity 
\cite{SeWi, HaKrLa1}.

In \cite{HaKrLa1} the exact non-commutative solitons for
the string field theories 
were obtained using the technique 
called solution generating technique, which is also useful for
the BPS case \cite{HaTe} \cite{Ha}.
For the brane-antibrane system, 
they assumed the form of action which does
not vanishes at the closed string vacuum $T=T_0$. 
Our action (\ref{Z2}), however, vanishes at the minimum $T=T_0=\infty$
and does not have the form assumed in \cite{HaKrLa1}.
Thus we should confirm whether 
their construction of the exact soliton
works for a noncommutative version of 
our brane-antibrane action (\ref{Z1}) or not.
In order to use the solution generating technique,
we regard the fields on the non-commutative field theory as 
operators on Fock space.
In \cite{HaKrLa1}
an almost gauge transformation was defined as 
\beqa
D_{\mu}^{(i)}  & \rightarrow & S^{(i)} D_{\mu}^{(i)} S^{(i) \dagger}, 
\nonumber \\
T & \rightarrow &  S^{(1)} T S^{(2) \dagger},
\eeqa
where $D^{(i)}$ is a covariant derivative operator
$D^{(i)}=d-i A^{(i)}$ and $i=1,2$.
Here 
$S^{(i)}$ is an almost unitary operator 
which satisfies
$S^{(i) \dagger} S^{(i)} =1 $ and 
$S^{(i)} S^{(i) \dagger}=1-P^{(i)}$
where $P^{(i)}$ is a projection operator.
First, we start from the trivial vacuum $A_\mu^{(i)}=0$
and $T=T_0(=\infty)$, which is a solution of the equations of motion
for the noncommutative version of the action (\ref{Z1}).
Then the configuration constructed by the above transformation 
becomes a nontrivial exact solution of the equations of motion
from the argument in \cite{HaKrLa1}.

We can see that the tension of this soliton is correct value.
The process of its calculation is almost same as in \cite{HaKrLa1}.
The only difference from \cite{HaKrLa1}
is the form of the action, especially the explicit form of terms of
field strengths without covariant derivatives.
These terms in our action are the sum of Born-Infeld actions
multiplied by the function of the tachyon, 
${\rm exp}(- \bar{T} T )$, or  
${\rm exp}(- T \bar{T} )$.
We note that 
the tachyon $\bar{T} T, T \bar{T}$ and
the field strength 
$\hat{F}_{\mu \nu}^{(i)}+\Phi_{\mu \nu} 
\sim [D_{\mu}^{(i)},D_{\nu}^{(i)}]$ 
are transformed to 
$|T_0|^2 (1-P^{(2)}), |T_0|^2 (1-P^{(1)})$ and 
$S^{(i)} [D_{\mu}^{(i)},D_{\nu}^{(i)}] S^{(i) \dagger} 
\sim \Phi_{\mu \nu} (1-P^{(i)})$
respectively.
Here we can obtain
\beq
V( \bar{T} T) [D_{\mu}^{(2)},D_{\nu}^{(2)}] =0,
\eeq
for the soliton configuration
from the equation 
$V( \bar{T} T) = 
V(\bar{T} T )-V(0) +V(0)= (V(\bar{T_0} T_0) -V(0)) (1-P^{(2)})  +V(0)
= V(0) P^{(2)}$
where $V(\bar{T} T) \sim {\rm exp} (- \bar{T} T )$ and
$V(\bar{T} T)=0$ at $T=T_0=\infty$.
Therefore for this soliton configuration constructed from the vacuum
where the action vanishes,
the sum of Born-Infeld actions remains to vanishes
except the gauge fields independent term, i.e. the tachyon potential.
Hence the soliton constructed in \cite{HaKrLa1}
is an exact solution of the brane-antibrane action which represents 
$N_1$ Dp-brane and $N_2$ anti Dp-brane, where 
$N_i={\rm dim (Ker(1-P^{(i)}))}$.

Note that the action evaluated at this soliton configuration has 
a non-zero value, while the action 
evaluated at the closed string vacuum $T=\infty$ vanishes. 
In \cite{HaKrLa1} they argued that
the action can not vanish even at the closed string vacuum
from an observation that 
the BPS brane after the tachyon condensation 
has non-zero mass.
In fact, our action has nonzero value for the soliton solution
and is consistent.

The properties of vanishing kinetic terms may
be required from the observation that at the closed string vacuum
in order to solve the $U(1)$ problem 
the strong coupling effects should be important and 
the vanishing kinetic terms almost mean the strong coupling 
physics \cite{Yi,sen22,BeHoYi,KlLaSh}.

Next we compute the brane-antibrane effective action as 
a sigma model partition function perturbatively up to $\alpha'^2$.
Hereafter we will restore the dimension-full parameter $\A'$ 
by including a factor $\A'/2$.
Here we use the regularized correlation function 
\ba
\langle \xi^\mu(\tau)\xi^\mu(\tau ') \rangle &=& \A' \!\!\! 
\sum_{m\in {\mb{Z} \ne 0}}
\f{1}{|m|}e^{im(\tau-\tau ')-\ep|m|}
=2 \A' 
\sum_{m\in {\mb{Z} > 0}} \frac{1}{m} {\rm cos} (m(\tau-\tau')) e^{-\ep|m|}
, \label{gb1}\\
\langle \psi^\mu(\tau) \psi^\mu(\tau ') \rangle &=&
-\f{i}{2}\sum_{r\in {\mb{Z}}+\f12}\f{r}{|r|}e^{ir(\tau-\tau')-\ep|r|}=
\sum_{m\in {\mb{Z}+\f12 > 0}} {\rm sin} (r(\tau-\tau')) e^{-\ep|r|}. 
\label{gf1}
\ea
This regularization keeps world-sheet supersymmetry and 
the spacetime gauge invariance which corresponds to 
world-sheet global symmetry \cite{Ts1, AnTs1}.
First we expand fields in (\ref{eq8}) as
\beqa
T \bar{T}(X) &=&
T \bar{T}(x) + \xi^{\mu} (D_\mu (T \bar{T}) )(x)+\cdots
+\frac{1}{4!} \xi^{\sigma} \xi^{\rho} \xi^{\nu} \xi^{\mu} 
(D_{\sigma} D_{\rho} D_{\nu} D_{\mu} (T \bar{T}) )(x)\cdots,
\nonumber \\
A^{(i)}_{\mu}(X) &=&
A^{(i)}_{\mu} (x) + \xi^{\nu} \pa_\nu A^{(i)}_\mu (x)+\cdots, \hspace{.5cm}
F^{(i)}_{\mu \nu}(X) = 
F^{(i)}_{\mu \nu}(x) + \xi^{\rho} (D_\rho F^{(i)}_{\mu \nu} )(x)+\cdots,
\nonumber\\
D_{\mu} T(X) &=&
D_{\mu} T(x) + \xi^{\nu} (\pa_\nu (D_{\mu} T) )(x)+ 
\frac{1}{2} \xi^{\rho} \xi^{\nu} (\pa_{\rho} \pa_\nu (D_{\mu} T) )(x)\cdots.
\eeqa
Then we can compute the partition function 
usually by the perturbation in $\alpha'$.
Since the actual computations are somewhat complicate,
we will only show the outline of the computation and the result below.

The gauge invariance of the effective action can be 
checked by replacing the $\pa_{\mu}$ by $D_{\mu}$ and 
picking the terms which depend on $A^{(i)}_{\mu}$.
For example, the coefficient of the term 
$A_{\rho}^{(-)} (\overline{D^{\mu} T} ) (\pa_{\rho} D_{\mu} T)$
is proportional to
\beq
\sum_{r,m>0} \frac{1}{m} 
\left( \frac{1}{r+m}+\frac{1}{r-m} \right) e^{-(r+m)\epsilon}
+\sum_{r,m>0} \frac{1}{r}
\left( \frac{1}{r+m}+\frac{1}{m-r} \right) e^{-(r+m)\epsilon}
- \sum_{r,m>0} \frac{2}{r m} e^{-(r+m)\epsilon},
\eeq
which is indeed zero, where $r \in {\bf Z}+\frac{1}{2}>0$ and
$m \in {\bf Z}>0$.
The other terms can be calculated explicitly by 
the formulae in \cite{AnTs1} except 
a finite constant 
\beq
\gamma_0={\rm \lim_{\epsilon \rightarrow 0}} \left(
\sum_{r,m>0} \frac{1}{m} 
\left( \frac{1}{r+m}+\frac{1}{r-m} \right) e^{-(r+m)\epsilon}
-(\log \epsilon)^2 \right).
\eeq
The result up to $\alpha'^2$ can be rearranged 
in a rather simple form:
\beqa
S  =Z&=&  T_9 e^{-\bar{T_{R}}T_{R} }
\left[ 2+8 \alpha' \log 2 D_{\mu}T_{R} \overline{D_{\mu} T_{R}} +
\alpha'^2 \pi^2 \left( \left( F^{(1)}_{\mu \nu R} \right)^2
+\left( F^{(2)}_{\mu \nu R} \right)^2 \right) 
\right. \nonumber \\
&& \hspace{2.0cm} +
4 \alpha'^2 \gamma_0 D_{\nu} D_{\mu} T_{R} D_{\nu} D_{\mu} \overline{T_{R}}
+32 \alpha'^2 i (\log 2)^2 F_{\mu \nu R}^{(-)} D_{\mu} T_{R} D_{\nu}
\overline{T_{R}}
\nonumber \\
&& \hspace{2.0cm} +
2 \alpha'^2 \left( 8 (\log2)^2 -\frac{1}{3} \pi^2 \right)
(D_{\mu} T_{R} D_{\mu} \overline{T_{R}})^2
- \alpha'^2 \frac{2}{3} \pi^2 (D_{\mu} T_{R})^2 ( D_{\nu} \overline{T_{R}})^2
\nonumber \\
&& \hspace{2.0cm}  +
\frac{\pi^2}{6}\alpha'^2 \left( (D_{\mu} D_{\nu}T_{R}) \overline {T_{R}}
+ T_{R} (D_{\mu} D_{\nu} \overline {T_{R}})   \right)
\nonumber \\
&& \hspace{2.0cm} \left. 
 \;\;\; \times \left( D_{\mu} D_{\nu} (\overline{T_{R}}T_{R}) 
+ D_{\nu} \overline{T_{R}} D_{\mu} T_{R} 
+ D_{\mu} \overline{T_{R}} D_{\nu} T_{R} 
\right)
\right], 
\label{a2}
\eeqa
where we renormalized the tachyon field only 
as 
\ba
\label{fa}
T&=&T_{R}+\alpha' \log \epsilon D_{\mu} D_{\mu} T_{R} 
+ \frac{1}{2} \alpha'^2  (\log \epsilon)^2 
D_{\nu} D_{\nu} D_{\mu} D_{\mu} T_{R}\nonumber\\
&&~~~~~~~~~~~~~~~~ + i \alpha^{\prime 2}
(\log \epsilon)^2 D_{\nu}F_{\nu\mu R}^{(-)}D_{\mu}T_{R},\nonumber\\
A_{\mu}^{(-)}&=&A_{\mu R}^{(-)}+\alpha^{\prime} \log \epsilon 
D_{\nu}F_{\nu\mu R}^{(-)}
\ea
Note that 
we obtain the form (\ref{a2}) 
by a field redefinition which corresponds to a renormalization
(\ref{fa}) from the two dimensional point of view.

\section{Conclusions and Future Directions}

In this paper we described the tachyon dynamics on the brane-antibrane 
by the method of the boundary string field theory. We constructed
the world-sheet boundary action of the brane-antibrane system
by using the boundary fermions. The remarkable point of 
this formalism is that for the special profile of the tachyon field the
calculation of the tachyon potential and the tensions of the lower
branes which are produced after the tachyon condensation can be performed
exactly. From these calculations we confirmed the `descent relation` of 
Non-BPS systems.

On the other hand in the case of including the all 
gauge fields it was
difficult to calculate the exact effective potential except RR
couplings. We found that the explicit form of the RR coupling in
boundary string field theory can be represented by the superconnection. 
Furthermore we also took the transverse scalars into account and showed
that Myers terms appear naturally.

We have also calculated several lower terms of the action of BSFT
in the $\alpha^{\prime}$ expansion. 
Further we discussed the general structure of the action and the result
was consistent with the arguments on noncommutative solitons. 
 
Our work raises some interesting questions and we hope to return to
these in future work.

\begin{itemize}
\item{As we have seen, the incorporation of one of the gauge field strength
$F^{(+)}$ on a brane-antibrane system can be treated as a
free theory and leads to the familiar noncommutative theory 
if the point-splitting regularization is employed. On the other hand,
the exact treatment of the other field strength $F^{(-)}$ is found to be
     difficult. 
Then for brane-antibrane systems 
it seem to be essential 
to ask whether we can
     express this effect as a sort of a noncommutative theory.}

\item{Our calculation here is performed assuming that the target space is
flat. Then it is natural to ask what will happen to the tachyon physics
if one considers a non-BPS D-brane system wrapping on a more complicated
manifold such as a Calabi-Yau manifold. In such a case one should take
world-sheet instantons into account. For example the discussion on their
RR couplings will be modified and 
couplings may be expressed by some ``stringy'' 
Chern characters. 
A related question is how much the world-sheet
supersymmetry has effects on the dynamics of tachyon condensation.}

\item{One more interesting question is the physical meaning of the Myers
terms which we have found for brane-antibrane systems and non-BPS
D-branes in the framework of BSFT.}
\end{itemize}

\vskip6mm
\noindent
{\bf Acknowledgements}

\vskip2mm
We would like to thank 
Y. Matsuo and K. Ohmori 
for useful discussions.
The works of S.T. and T.T.
were supported in part by JSPS Research Fellowships for Young 
Scientists. \\

\noindent
{\bf Note added}: 

While preparing this paper for publication,
we received the preprint \cite{Ho}
in which the world-sheet action 
for brane-antibrane system was given
and the preprint \cite{KrLa}
in which the BSFT for brane-antibrane system was discussed.
%which substantially overlaps the present work.

\appendix
\setcounter{equation}{0}
\section{Notation for Spinors and Some Formulas}

Here we summarize our notation for spinors following \cite{GaMy} and
after that  we show some calculations which is needed for the derivation
of the RR coupling in section 4.

The ten dimensional $\Gamma$-matrices $\Gamma_A\ ^B,\
( A,B=1,2,\ddd,32)$ are defined by the following Clifford algebra:
\ba
 \{ \Gamma^\mu,\Gamma^\nu \}=-2\eta^{\mu\nu}, \ \ \
\eta^{\mu\nu}=\mbox{diag}(-1,1,\ddd,1).
\ea 
Note that we distinguish spinor and adjoint spinor indices as subscripts 
and superscripts, respectively. The charge conjugation matrix $C^{AB},
C^{-1}_{AB}$, which obey the relations:
\ba
C\Gamma^\mu C^{-1}=-(\Gamma^\mu)^{T},
\ea
can raise or lower these spinor indices. Therefore we can omit the matrix $C$ as\ba
(\Gamma^\mu)_{AB}=C^{-1}_{BC}(\Gamma^{\mu})_A\ ^C, \ \ \
(\Gamma^\mu)^{AB}
=C^{AC}(\Gamma^{\mu})_C\ ^B.
\ea
Note also from the above equations it is easy to see
\ba
C_{AB}=-C_{BA},\ \ \ (\Gamma^\mu)_{AB}=(\Gamma^\mu)_{BA}.
\ea
We also define $(\Gamma_{11})_A\ ^B$ as
\ba
\Gamma_{11}=\Gamma_{0}\Gamma_{1}\ddd \Gamma_{9},
\ea
and the chirality projection matrix $P_{\pm}$ is defined by
\ba
P_{\pm}=\f{1}{2}(1\pm \Gamma_{11}).
\ea
The matrix $\Gamma_{11}$ satisfies the following identities
\ba
(\Gamma_{11})_{AB}=(\Gamma_{11})_{BA}, \ \ (\Gamma_{11})^2=1, \ \ \{\Gamma_{11},\Gamma^{\mu}\}=0.
\ea

Before we will discuss the calculation of the RR couplings, let us now show some useful formulae. The first one is about the trace of $\Gamma$-matrices:
\ba
\mbox{Tr}[\Gamma^{\mu_0\mu_1\ddd\mu_p}\Gamma^{01\ddd p}]=
32\ (-1)^{\f{p(p-1)}{2}}\ \ep^{\mu_0\mu_1\ddd\mu_p}\ \ (0\le\mu_i\le p),
\label{trace1}
\ea
where
$\Gamma^{\mu_0\mu_1\ddd\mu_p}=
1/p! \, 
(\Gamma^{\mu_0}\Gamma^{\mu_1}\dd\Gamma^{\mu_p}-\Gamma^{\mu_1}\Gamma^{\mu_0}\dd
\Gamma^{\mu_p}+\ddd )$ denotes the antisymmetrized $\Gamma$-matrices.
The second one is the famous relation between the $\Gamma$-matrices 
and the differential forms. More explicitly, a $r$-form in ten dimension:
\ba
C=\f{1}{r!}C_{\mu_1\mu_2\ddd\mu_r}dx^{\mu_1}dx^{\mu_2}\ddd dx^{\mu_r},\label{c1}
\ea
corresponds to the following $32\times 32$ matrix:
\ba
\hat{C}=\f{1}{r!}C_{\mu_1\mu_2\ddd\mu_r}\Gamma^{\mu_1\mu_2\ddd\mu_r}.\label{c2}
\ea
This correspondence preserves the multiplication as 
\ba
:\hat{C_1}\hat{C_2}:=\f{1}{(r_1+r_2)!}(C_1\wedge
C_2)_{\mu_1\mu_2\ddd\mu_{(r_1+r_2)}}\Gamma^{\mu_1\mu_2\ddd\mu_{(r_1+r_2)}},
\label{c3}
\ea
where $:\ :$ denotes the antisymmetrization.

Let us now turn to the derivation of the RR couplings. It involves the
computations of the correlation functions on a disk 
whose boundary is on a D$p$-brane. 
We assume its world-volume extends in the
direction $x^0,x^1,\ddd,x^p$. 
Then it is easier to calculate the
correlation functions 
by performing T-duality transformation.  
This transformation is given with
respect to the spin operators by
\ba
S_A\ \to\ S_A \ &,&\ \ti{S}_A\ \to\ M_{A}\ ^{B}\ti{S}_B, \no
M_{A}\ ^{B}&=&\left\{\begin{array}{c}
\pm i\Gamma^{0}\Gamma^{1}\ddd\Gamma^{p}\ \ \ (p=\mbox{even})    \\
\pm \Gamma^{0}\Gamma^{1}\ddd\Gamma^{p}\Gamma_{11}\ \ \ (p=\mbox{odd})
\end{array} \right .,
\ea
where $S_A$ and $\ti{S}_B$ 
denote the left-moving and right-moving spin operators, respectively; the sign ambiguity $\pm$ depends on the conventions and we choose the the plus sign.
 The above rule can be derived by requiring that the OPE's of the
left-moving fermions $\psi^{\mu}$ and spin operators $S_A$ have the same
structure as those of right-moving ones $\ti{\psi}^\mu, \ \ti{S}_B$
after one performs the T-duality transformation. Using these facts, the
RR couplings on both a non-BPS D-brane and a brane-antibrane system are summarized as the following form up to the overall normalization (see eq.(\ref{rr1}) and (\ref{rr2})):
\ba
S=\sum_{r=0}^{p+1}\f{1}{r!}K_{\mu_1\mu_2\ddd\mu_r}\ \mbox{Tr}[\ P_{-}\hat{C}M\Gamma^{\mu_1\mu_2\ddd\mu_r}\ ],
\ea  
where $K_{\mu_1\mu_2\ddd\mu_r}\ \ (0\le \mu_i \le p)$ is a $r$-form which depends on the field-strength
 and the tachyon field; $\hat{C}$ denotes the RR-sector vertex as
defined in eq.(\ref{rr0}). If one takes the transverse scalars into
account, one should also discuss $K_{\mu_1\mu_2\ddd\mu_r}$ for $\mu_i
\ge p$. However such a case can also be treated similarly and we omit this.
Then we can write down the RR couplings 
explicitly up to the overall normalization which is independent of $r$ as follows:
\ba
S=\sum_{q,r=0}^{p+1}\f{1}{q!r!}\ \delta_{p+1,q+r}\ \ep^{\mu_1\ddd\mu_q\nu_1\ddd\nu_r}C_{\mu_1\ddd\mu_q}K_{\nu_1\nu_2\ddd\nu_r},
\ea
where we have used the formula (\ref{trace1}). Finally these couplings are 
written in the language of the differential forms as
\ba
S=\int_{\Sigma_{(p+1)}} C\wedge K,
\ea
where $\Sigma_{(p+1)}$ denotes the world-volume of the $p$-brane.

\end{document}